\documentclass[pre,twocolumn]{revtex4-1}

\usepackage{graphicx}
\usepackage{dcolumn}
\usepackage{bm}

\usepackage[utf8]{inputenc}
\usepackage[T1]{fontenc}
\usepackage{mathptmx}
\usepackage{bm}
\usepackage{pdfrender}

\usepackage{amsmath}
\usepackage{mathtools}
\usepackage{amssymb}
\usepackage{tikz}

\usepackage{pst-node}
\usepackage{verbatim}   
\usepackage{color}      
\usepackage{subfigure}  
\usepackage{hyperref}   

\usepackage{braket} 

\newcommand{\bnabla}{\mbox{\boldmath $\nabla$}}

\newcommand{\ba}{\begin{eqnarray}}
\newcommand{\ea}{\end{eqnarray}}
\newcommand{\be}{\begin{equation}}
\newcommand{\ee}{\end{equation}}

\begin{document}

\title{Run-and-tumble oscillator:  moment analysis of stationary distributions} 

\author{Derek Frydel}%
 \email{dfrydel@gmail.com}
\affiliation{ 
Department of Chemistry, Universidad Técnica Federico Santa María, Campus San Joaquin, 7820275, Santiago, Chile
}%


\date{\today}

\begin{abstract}


When it comes to active particles, even an ideal-gas model in a harmonic potential poses a mathematical challenge.  
An exception is a run-and-tumble model (RTP) in one-dimension for which a stationary distribution is known exactly.  
The case of two-dimensions is more complex but the solution is possible.  Incidentally, in both dimensions the stationary 
distributions correspond to a beta function.  
In three-dimensions, a stationary distribution is not known but simulations indicate that it does not have a beta 
function form.  
The current work focuses on the three-dimensional RTP model in a harmonic trap.    
{The main result of this study is the derivation of the recurrence relation for generating moments of a 
stationary distribution.}
These moments are then used to recover a stationary distribution using the Fourier-Lagrange expansion.   
\end{abstract}

\pacs{
}

\maketitle

\section{Introduction}

Ideal-gas of active particles in a harmonic trap at first glance appears like a simple toy model with ready 
solutions and useful insights.  The fact that such solutions are still lacking, or in the making, highlights
the fact that active matter, even at the most basic level, is a challenge and experimentation with alternative 
formulations is needed and justified.

In this work we focus on stationary marginal distributions, that we designate as $p$, of run-and-tumble 
particles (RTP) in a 
harmonic trap.  In one- \cite{Tailleur08,Tailleur09,Dhar19,Basu20} and two-dimensions \cite{Frydel22c}, 
those distributions have a beta functional form.  In three-dimensions, no exact expression for a distribution 
is available.  In this work, instead of obtaining an expression for $p$ directly, we 
calculate moments of that distribution.  The moments are generated by 
the recurrence relation obtained by transforming the Fokker-Planck equation.  
A stationary distribution $p$ is then recovered from those moments using the Fourier-Legendre 
series expansion.

The main analysis in this work is carried out for a system at zero temperature and for a harmonic potential in a 
single direction $u = \frac{1}{2} K x^2$ (embedded in a higher dimension).   This 
makes the analysis simpler since the system is effectively one-dimensional.  To extend the results to finite temperatures, 
we use the convolution construction \cite{Frydel22c,Frydel23}, which is equivalent to 
Gaussian smearing of a distribution at zero temperature.  
It also turns out that the moments of a stationary distribution for a potential $u = \frac{1}{2} K x^2$
can be related to the moments of a stationary distribution for 
an isotropic potential $u = \frac{1}{2} K r^2$.  This permits us to extend our analysis 
in s straightforward way to isotropic potentials.  

To place the current work in a larger context, we mention a number of previous contributions to active particles 
in a harmonic potential.   An extension of the RTP model in 1D and zero temperature to three discrete swimming 
velocities was considered in \cite{Basu20}.  The RTP model in two-dimensions (2D) with four discrete swimming 
velocities was investigated in \cite{Scher22}.  A stationary distribution of active Brownian particles in 2D and for 
a finite temperature represented as a series expansion was considered in \cite{Dhar20}.   Dynamics of active Brownian particles (ABP) 
was recently investigated in \cite{Cargalio22}.  {A unifying approach to theoretical treatment of ABP and AOUP models 
in a harmonic trap was carefully investigated in \cite{Carpini22}}.  Rare events in the context of active particles in a 
harmonic potential were considered in \cite{Farago22}.  Active particles in a harmonic chains was considered in 
\cite{Gupta21,Kundu21,Basu22}.  Experimental realizations of active particles in a harmonic trap are found in 
\cite{Brady16}, using acoustic traps, and in \cite{Lowen22}, using optical tweezers.  Entropy production rate of 
active particles in a harmonic trap was considered in \cite{Dabelow19,Caprini19,Dabelow21,Pruessner21,Frydel22a,Frydel23}. 

{As an exact analysis of the RTP and ABP particles in a harmonic trap can be challenging, the active Ornstein-Uhlenbeck
particles (AOUP) is more straightforward.  The AOUP model has been developed to capture a behavior of a passive particle 
in a bath of active particles \cite{Maggi14,Szamel14}.  Stationary distributions of this model in a harmonic trap have a Gaussian 
functional form, the same as that for passive Brownian particles, but with 
an effective temperature.  Theoretical aspects of the AOUP model have been extensively investigated in \cite{Martin21,Caprini21}.}

This paper is organized as follows.  In Sec. (\ref{sec:sec1}) we consider RTP particles in a harmonic 
trap in 1D.  We consider distributions in a position and a velocity space to identify the presence of "nearly immobile" 
particles.  
In Sec. (\ref{sec:sec2}) we consider the RTP particles in a harmonic trap $u=\frac{1}{2}Kx^2$ embedded in 2D.  
In Sec. (\ref{sec:sec3}) we consider RTP particles embedded in 3D.   By transforming the Fokker-Planck equation, 
we obtain a recurrence relation for generating moments of the stationary distribution.  From the moments we then 
recover distributions using the Fourier-Legendre expansion.  
In Sec. (\ref{sec:sec-TF1D}) we extend the previous results to finite temperatures then in Sec. (\ref{sec:sec4}) 
to isotropic harmonic potentials.  
In Sec. (\ref{sec:summary}) we summarize the work and provide concluding remarks.

\section{RTP particles in 1D}
\label{sec:sec1}

We start with the simple case: RTP particles in a harmonic trap $u = \frac{1}{2} Kx^2$ in 1D  
\cite{Tailleur08,Tailleur09,Dhar19,Basu20,Frydel22c}.  
{Apart for looking into stationary distributions in a velocity space, Sec. (\ref{sec:sec1A}), the section reviews previously
derived results.}

In one-dimension, swimming orientations are limited to two values, $v_{swim}=\mp v_0$ and 
the Fokker-Planck formulation yields two coupled differential equations \cite{Frydel22c}: 
\ba
&&\dot p_+   = \frac{\partial}{\partial x}  \left[ \left(\mu K x  - v_0\right) p_+ \right]      +     \frac{ 1}{2\tau} \left( p_-   -   p_+  \right)
\nonumber\\
&& \dot p_-    =  \frac{\partial}{\partial x}  \left[ \left(\mu K x  + v_0\right) p_- \right]    +     \frac{1}{2\tau}  \left( p_+   -     p_-  \right), 
\ea
where $p_+$ and $p_-$ are the distributions of particles with forward and backward direction, $\tau$ is the 
persistence time (that determines the average time a particle persists in a given direction), 
and $\mu$ is the mobility.  No thermal fluctuations are taken into account.  

The two equations in a stationary state $\dot p_{\pm}=0$ and dimensionless units become
\ba
&&
0  = \frac{\partial}{\partial z}  \left[ \left(z  - 1\right) p_+ \right]      +     \frac{ \alpha}{2} \left( p_-   -   p_+  \right)
\nonumber\\
&& 
0 =  \frac{\partial}{\partial z}  \left[ \left(z  + 1\right) p_- \right]    +     \frac{ \alpha}{2}  \left( p_+   -     p_-  \right).  
\label{eq:FP1D}
\ea
where
$$
z  =  \frac{\mu K x} {v_0}, 
$$ 
is dimensionless distance and 
\be
\alpha = \frac{1}{\tau \mu K},
\label{eq:alpha}
\ee
is the dimensionless rate of an orientational change.  Note that in one-dimension the new direction of motion 
is chosen at the rate $\alpha/2$ (rather than $\alpha$).  The reason for this is that at an instance that a particle 
changes its direction, there is $1/2$ probability it will select the same orientation.  This problem does not arise 
for higher dimensions and $\alpha$ is the actual rate at which a particle changes its orientation.  
This should be kept in mind when we later compare the results for different dimensions.

The two coupled equations in Eq. (\ref{eq:FP1D}) can be combined into a single differential equation for the total distribution 
$p=p_+ + p_-$, 
\be
0 = (2-\alpha)zp    -   (1-z^2)p', 
\label{eq:FP-p-1D}
\ee
which when solved yields \cite{Tailleur08,Tailleur09,Dhar19,Basu20,Frydel22c} 
\be
p = A (1-z^2)^{\frac{\alpha}{2}-1},
\label{eq:pz-1d}
\ee
and where the normalization factor, that assures $\int_{-1}^{1}dz\, p(z)=1$, is given by
\be
A = \frac{\Gamma \left(\frac{\alpha}{2} + \frac{1}{2}\right)}{\sqrt{\pi } \Gamma \left(\frac{\alpha}{2}\right)}.  
\ee
Note that in the absence of thermal fluctuations $p$ is defined on $[-1,1]$ as a result of 
a swimming velocity having a fixed magnitude, which restricts how far a particle can move away from 
a trap center.  

The distribution $p$ in Eq. (\ref{eq:pz-1d}) can be either concave, with a majority of particles accumulated at the 
trap borders as a result of slow orientational change, or convex, with a majority of particles concentrated around 
a trap center as a result of fast orientational change.  The crossover between the 
two behaviors occurs at $\alpha=2$, at which point $p$ is uniform on the interval $[-1,1]$.

In addition to $p$, it is possible to obtain distributions for a specific swimming direction:  
\be
p_{\pm } = \frac{A}{2} \left(1 \pm z \right)  \left( 1 - z^2 \right)^{ \frac{\alpha}{2} - 1 }.  
\label{eq:ppm}
\ee 
The expression above can be verified if inserted into Eq. (\ref{eq:FP1D}).

\subsection{distribution in $w$-space}
\label{sec:sec1A}

For slow rates of orientational change, that is, for $\alpha<2$, the accumulation of particles 
near the trap border takes the form of a divergence at $z=\pm 1$, see Eq. (\ref{eq:pz-1d}). 
That divergence can be linked to the presence of "nearly immobile" particles accumulated 
at the trap border.  

The existence of "nearly immobile" particles can be verified from a velocity distribution, 
manifested as a divergence at $v=0$. In the overdamped regime, the two contributions to 
a velocity are a swimming velocity plus a contribution of a linear force of a harmonic trap,   
$
v = -\mu K x \pm  v_0, 
$ 
in the dimensionless units given by 
\be
w = -z \pm 1, 
\label{eq:w}
\ee
where $w = v/ v_0$ is the dimensionless velocity.   

A distribution in $w$-space can be inferred from a positional distribution in Eq. (\ref{eq:ppm}) by 
applying the change of variables suggested by Eq. (\ref{eq:w}).   For particles with forward orientation, 
the substitution $z = -w + 1$ into $p_+(z)$ leads to 
$$
p_w^+ 
= \frac{A}{2} w^{\frac{\alpha}{2}-1} (2 - w)^{\frac{\alpha}{2}}, ~~~ \text{defined on ~~} 0 < w < 2.  
$$
The reason why the distribution for the forward swimming velocity is defined on $[0,2]$ can be understood from 
Eq. (\ref{eq:w}) and the fact that $z$ is defined on $[-1,1]$.  Given that $p_w^-(w) = p_w^+(-w)$, 
a complete distribution defined on $[-2,2]$ is 
\be
p_w = \frac{A}{2} |w|^{\frac{\alpha}{2}-1} (2-|w|)^{\frac{\alpha}{2}}.  
\label{eq:pw-1d}
\ee
The divergence at $w=0$ signals the presence of "nearly immobile" particles.  
We characterize these particles as "nearly immobile" since 
 $\lim_{\varepsilon\to 0}\int_{-\varepsilon}^{\varepsilon} dw\, p_w = 0$, which implies that 
there are no particles with zero velocity.  Rather, there are particles 
whose velocity slowly converges to zero without ever attaining it.

Comparing Eq. (\ref{eq:pw-1d}) with Eq. (\ref{eq:pz-1d}) confirms that divergences in both distributions
appear/disappear for the same value of $\alpha$.  This coextensiveness implies that 
the "nearly immobile" particles are concentrated around trap borders at $z=\pm 1$.  Only 
at $z=\pm 1$ a particle can attain zero velocity, and since 
$\lim_{\varepsilon\to 0}\int_{1-\varepsilon}^{1} dz\, p(z) = \lim_{\varepsilon\to 0}\int_{-1}^{-1+\varepsilon} dz\, p(z) =0$, 
no particle can reach this position, except for $\alpha=0$.

\begin{figure}[hhhh] 
 \begin{center}
 \begin{tabular}{rrrr}
\includegraphics[height=0.19\textwidth,width=0.24\textwidth]{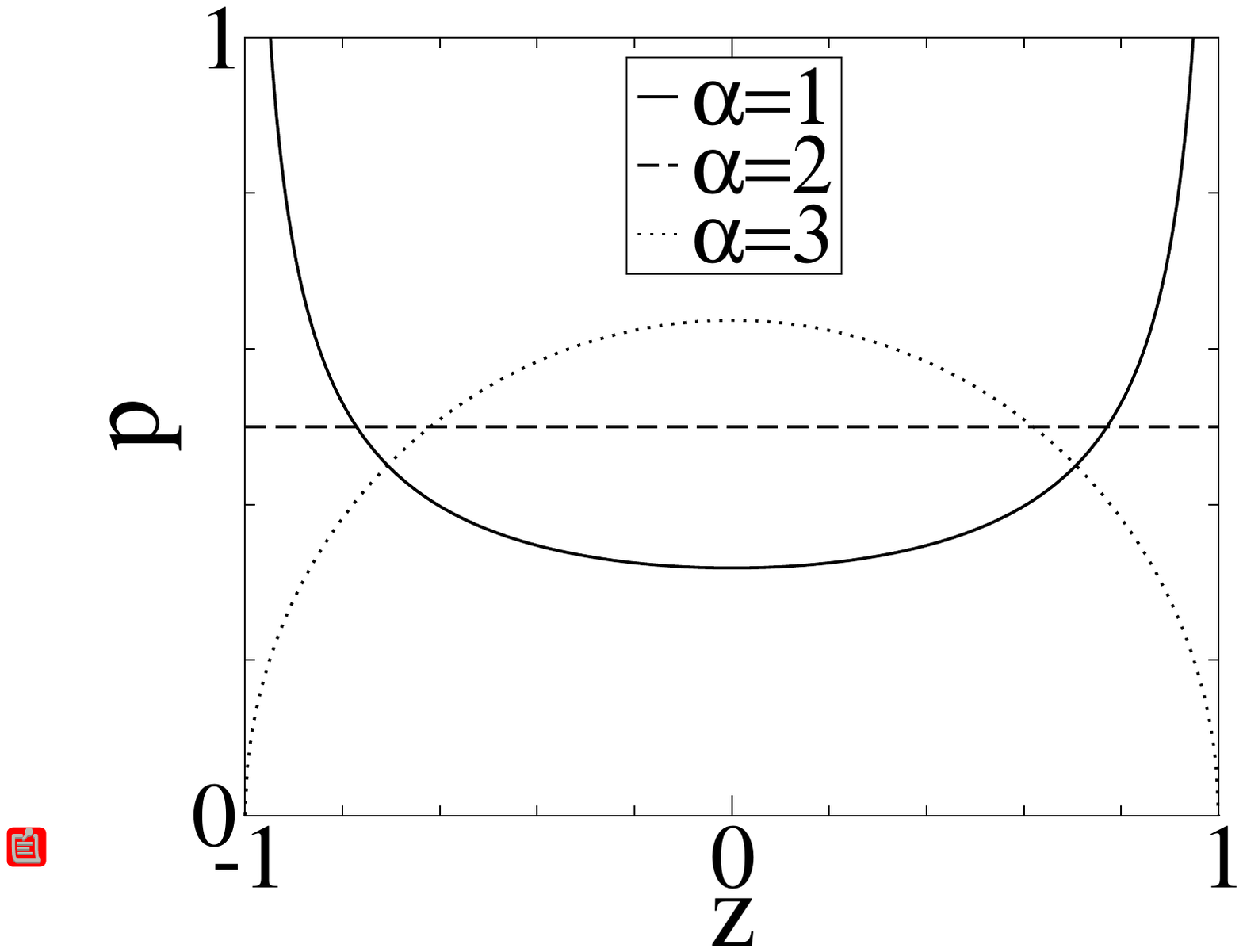}  
\includegraphics[height=0.19\textwidth,width=0.24\textwidth]{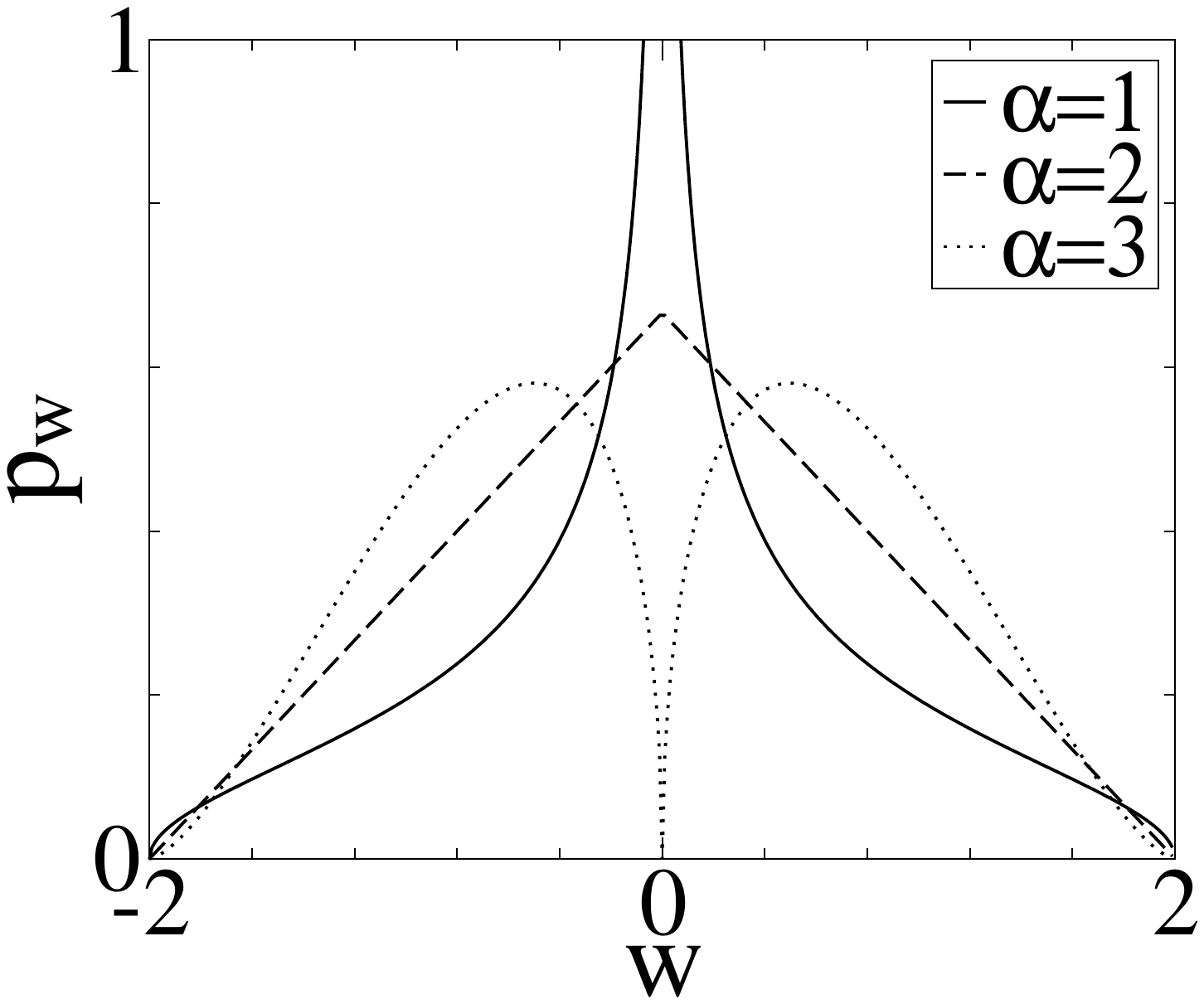}   
 \end{tabular}
 \end{center} 
\caption{ Stationary distributions in $z$- and $w$-space for different values of $\alpha$.  Divergence
emerges at the crossover $\alpha=2$ and is linked to the presence of immobile particles at  
trap borders.  }
\label{fig:fig1} 
\end{figure}

In Fig. (\ref{fig:fig1}) we plot $p$ and $p_w$ for three values of $\alpha$.  
\graphicspath{{figures/}}
At the crossover, $\alpha=2$, both distributions reduce to simple shapes:  
$p$ is flat and $p_w$ is triangular.  
Then for $\alpha<2$, both distributions develop divergences.

\subsection{moment analysis}

In this section, we
briefly analyze even moments $\langle z^{2n} \rangle$ of a distribution $p(z)$ in Eq. (\ref{eq:pz-1d}).  
(Odd moments are zero due to even symmetry of $p$).  
The moments can be calculated directly from $p$ using $\langle z^{2n}\rangle = \int_{-1}^1dz\, p(z) z^{2n}$:   
\be
\langle z^{2n}\rangle = \frac{ \Gamma\left( n + \frac{1}{2}  \right)  \Gamma\left( \frac{1}{2}  +   \frac{\alpha}{2} \right) } { \sqrt{\pi} \, \Gamma\left( \frac{1}{2}  +  \frac{\alpha}{2}  +  n \right) },
~~~~ \text{for } n=1,2,\dots
\label{eq:z2n-1d}
\ee
The moments are monotonically decreasing with increasing $n$.  The infinite sum $\sum_{n=0}^{\infty} \langle z^{2n}\rangle$ 
can be evaluated exactly and is given by 
\be
\sum_{n=0}^{\infty} \langle z^{2n}\rangle  = 
  \begin{cases}
      \frac{\alpha-1}{\alpha-2} & \text{if $\alpha>2$}\\
    \infty  & \text{if $\alpha\leq 2$}, 
  \end{cases}
\label{eq:sum-1dl}
\ee
where at the crossover the sum is seen to diverge.  By representing the infinite sum by its generating 
function, we can connect this behavior to the divergence in $p$:   
$$
\sum_{n=0}^{\infty} \langle z^{2n} \rangle = \left\langle \frac{1}{1-z^2} \right\rangle.  
$$
As particles accumulate at the borders, $z\to \pm 1$ and the above expression diverges.  
This explains the divergence of the sum in Eq. (\ref{eq:sum-1dl}).

\section{RTP oscillator in 2D}
\label{sec:sec2}

We next consider an RTP oscillator with 1D geometry, $u=\frac{1}{2} Kx^2$, but embedded in 2D space.  
To set up the problem, 
we start with a Fokker-Planck equation for RTP particles in an arbitrary confinement:
\be
\dot\rho  =  -\bnabla \cdot \left[ \left( \mu {\bf F} + v_0 {\bf n} \right) \rho \right]   +  \hat L\, \rho, 
\label{eq:FP-RTP}
\ee
where ${\bf n}$ is the unit vector designating an orientation of the swimming velocity $v_{swim} = v_0 {\bf n}$, 
in 2D defined as ${\bf n} = (\cos\theta,\sin\theta)$, where $\theta$ is the angle of the orientation.  
The evolution of ${\bf n}$ is governed by the operator $\hat L$ given by 
$$
\hat L\, \rho = \frac{1}{\tau} \left[ -\rho + \frac{1}{2\pi} \int_0^{2\pi} d\theta\, \rho(x,\theta) \right].  
$$
The two terms imply that particles with a given orientation $\theta$ vanish with the rate $\tau^{-1}$ and 
reappear with the same rate at another location that depends on the marginal distribution
$$
p =  \int_0^{2\pi} d\theta\, \rho(x,\theta).  
$$  
Note that $\int_0^{2\pi} d\theta\,\hat L\, \rho = 0$.  This 
condition is necessary if the total number of particles is to be conserved.

For $u=\frac{1}{2} Kx^2$, the external force is ${\bf F} = -Kx \, {\bf e}_x$.  In a steady-state, only 
the component of ${\bf v}_{swim}$ in the $x$ direction is relevant, ${\bf v}_{swim} \cdot {\bf e}_x = v_0 \cos\theta$.
This results in an effectively one-dimensional system governed by the following stationary Fokker-Planck equation: 
\be
0  =  \frac{\partial}{\partial z} \left[ \left( z  -  \cos\theta \right) \rho \right]  
-  \alpha\rho      +     \frac{\alpha}{2\pi}   p,
\label{eq:FP-LH}
\ee
given in dimensionless units.  

The above Fokker-Planck equation can be interpreted as representing RTP model in 1D with 
continuous distribution of velocities, and what constitutes a generalized RTP model \cite{Frydel21b}. 
For a truly 1D RTP model, the distribution of swimming velocities is 
$P \propto \delta(v_{swim}+v_0) + \delta(v_{swim}-v_0)$.  
The Fokker-Planck equation in Eq. (\ref{eq:FP-LH}) represents a system for the following distribution of 
swimming velocoties:  
$P \propto 1/\sqrt{v_0^2 - v_{swim}^2}$.  See Appendix A in \cite{Frydel21a}.    

{There is no straightforward procedure to reduce Eq. (14) to a differential equation 
for $p$ but it is possible to infer it from the moments of $p$.   (How to calculate such moments will
be demonstrated when we analyze a system in 3D).  Because a moment formula in 2D was 
determined to have a similar structure to that in 1D, it was, in turn, possible to infer 
that a differential equation for $p$ should have the same structure as that for a system in 1D. 
See Eq. (\ref{eq:FP-p-1D}).  
For 2D, the differential equation for $p$ was determined to be \cite{Frydel22c} }
\be
0 = (1 - 2\alpha)zp    -   (1-z^2)p',
\ee
where the solution is a beta distribution 
\be
p(z) = A (1-z^2)^{\alpha - \frac{1}{2}}, 
\label{eq:pz-2dl}
\ee
and where the normalization constant is given by 
\be
A = \frac{\Gamma \left(\alpha + 1\right)} { \sqrt{\pi } \Gamma \left(\alpha + \frac{1}{2} \right)}.  
\ee

\subsection{distribution in $w$-space}
\label{sec:sec2A}

For a system embedded in 2D, the velocity component in the $x$-direction is 
$
v = -\mu K x + v_0 \cos\theta, 
$
in reduced units given by 
\be
w = -z + \cos\theta.  
\label{eq:w2}
\ee
Compare this with Eq. (\ref{eq:w}).

The distribution $p_w$ can be obtained from Eq. (\ref{eq:FP-LH}) by substituting for $p$ with expression given in 
Eq. (\ref{eq:pz-2dl}), followed by the change of variables $z=-w+\cos\theta$, followed by the integration over all 
orientations.  This yields an inhomogeneous first order differential equation 
\be
(1 -  \alpha) p_w    +   w p_w'     =   - \frac{\alpha }{2\pi}  I, 
\label{eq:pw2DL}
\ee
where 
\be
I(w) = 2 A \int_{-1}^{1-w} ds \,  \left[1 -  s^2  \right]^{\alpha - \frac{1}{2}}  \left[ 1 - (s+w)^2 \right]^{-\frac{1}{2}}.  
\ee
The solution is a distribution defined on $[-2,2]$
\be
p_w   =   \frac{\alpha }{2\pi}  |w|^{\alpha-1}  \left[ \int_{|w|}^{2} dw'\,  w'^{-\alpha} I(w')\right].  
\label{eq:pw-2dl}
\ee

Although $p_w$ has a more complicated form compared to that in Eq. (\ref{eq:pw-1d}) for a system in 1D,
its general structure remains similar.  Divergence at $w=0$ comes from the factor $|w|^{\alpha-1}$, which 
signals the existence of nearly immobile particles for $\alpha>1$ and suggests the 
crossover at $\alpha=1$.  This, however, is not corroborated by $p$ in Eq. (\ref{eq:pz-2dl}), 
in which case divergences at the trap border emerge for $\alpha>1/2$.  

The reason why divergences in $p$ disappear at a lower value of $\alpha$ 
is a result of averaging procedure used to obtain $p$ from $\rho$.   
Even if a distribution $\rho$ exhibits divergences up to $\alpha=1$,
the averaging procedure $p = \int_{0}^{\pi} d\theta\,\rho$ smooths those divergences and effectively
makes them disappear for $\alpha<1/2$.  
In Appendix (\ref{sec:app0}) we analyze distributions $\rho$ in more detail to back up these claims.

\begin{figure}[hhhh] 
 \begin{center}
 \begin{tabular}{rrrr}
\hspace{-0.3cm}\includegraphics[height=0.16\textwidth,width=0.17\textwidth]{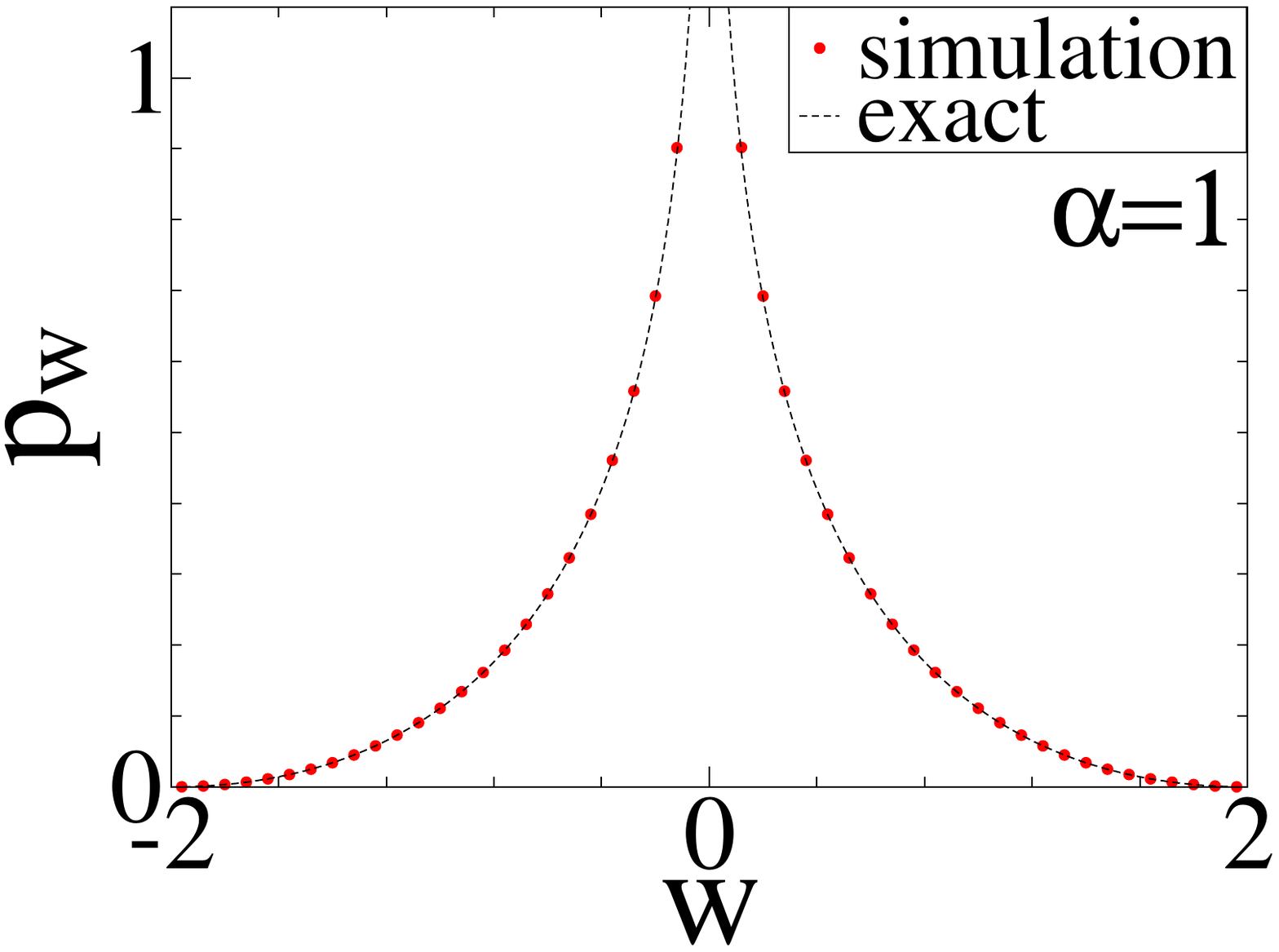} &
\hspace{-0.2cm}\includegraphics[height=0.16\textwidth,width=0.17\textwidth]{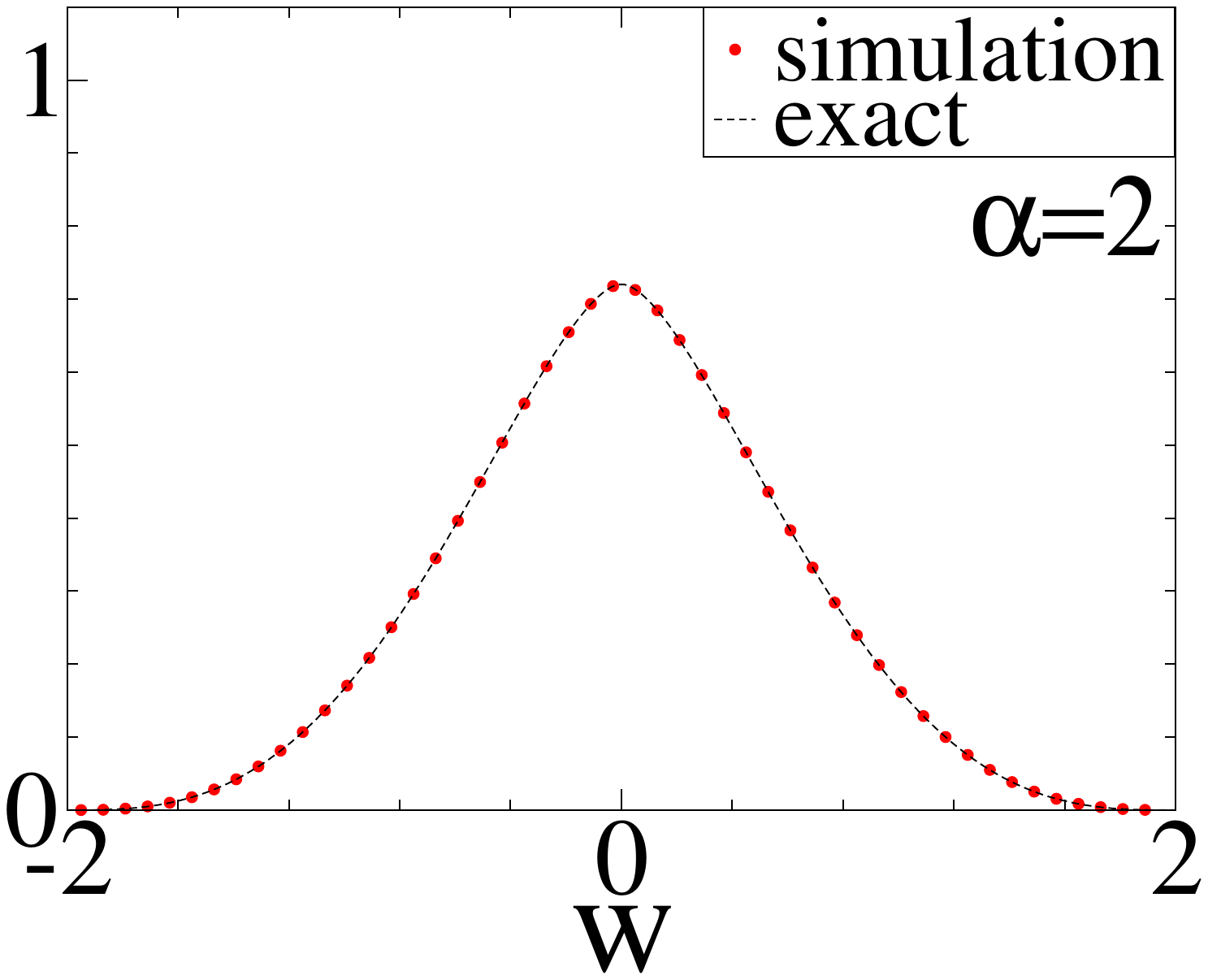} &
\hspace{-0.2cm}\includegraphics[height=0.16\textwidth,width=0.17\textwidth]{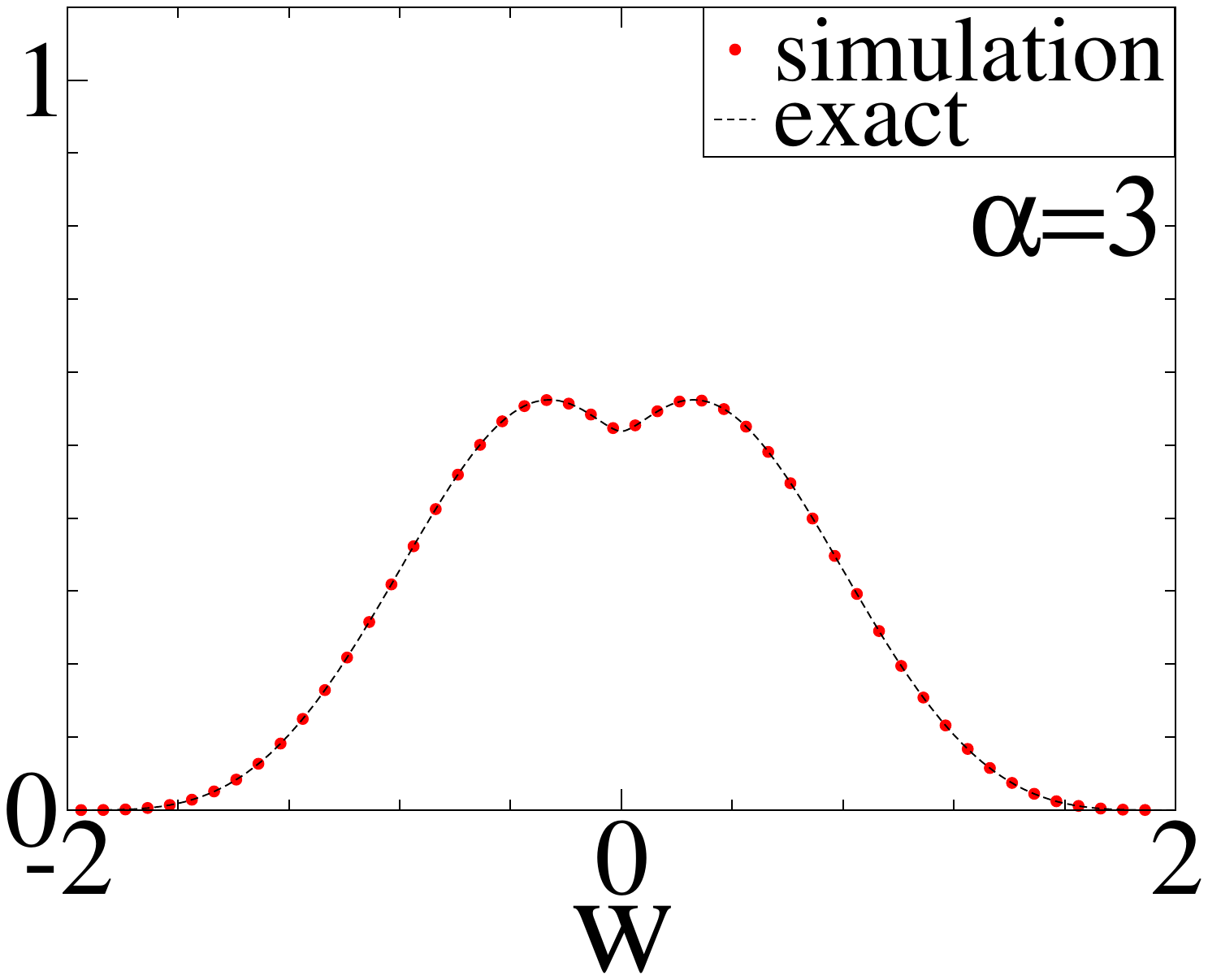} \\
\hspace{-0.3cm}\includegraphics[height=0.16\textwidth,width=0.17\textwidth]{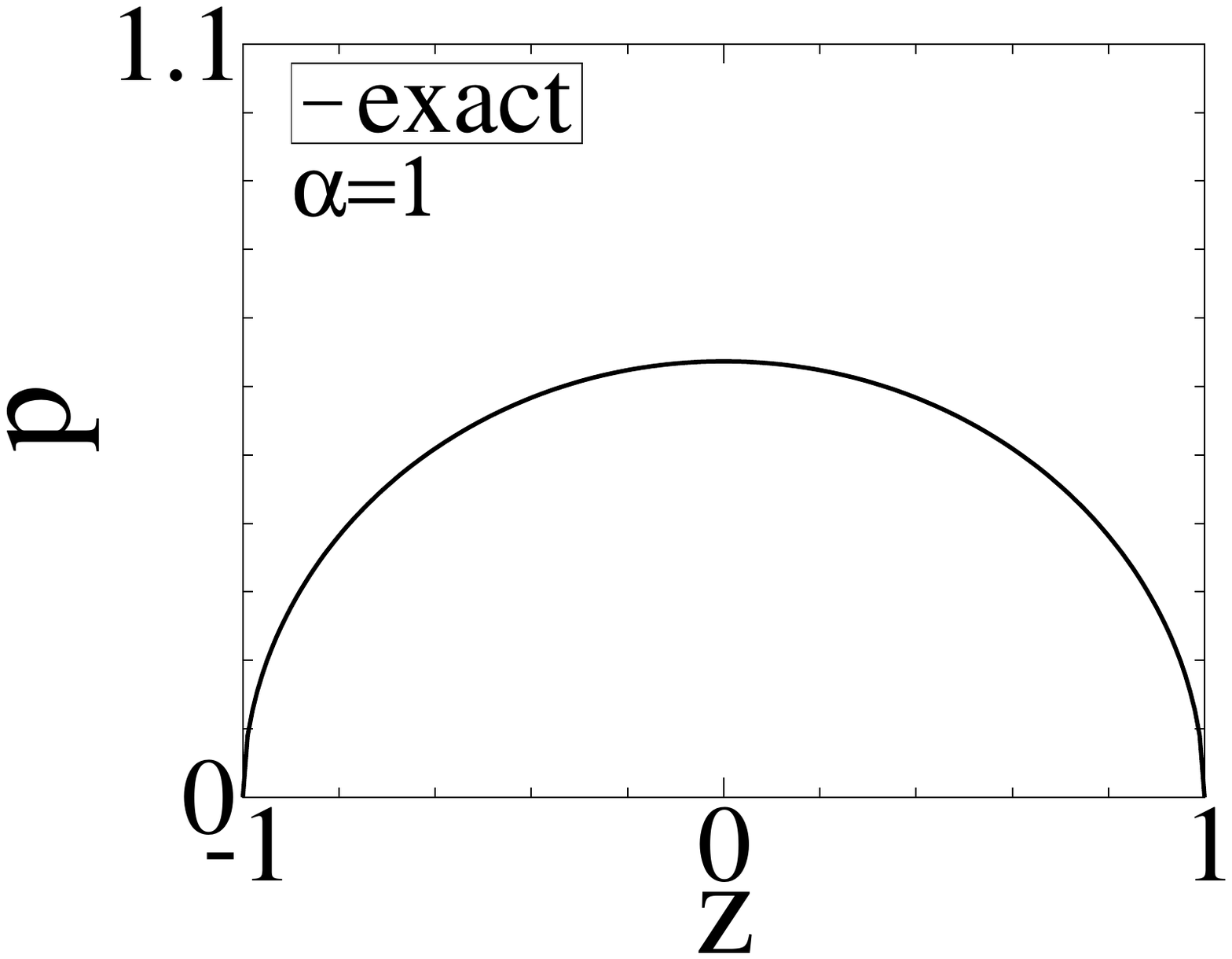} &
\hspace{-0.2cm}\includegraphics[height=0.16\textwidth,width=0.17\textwidth]{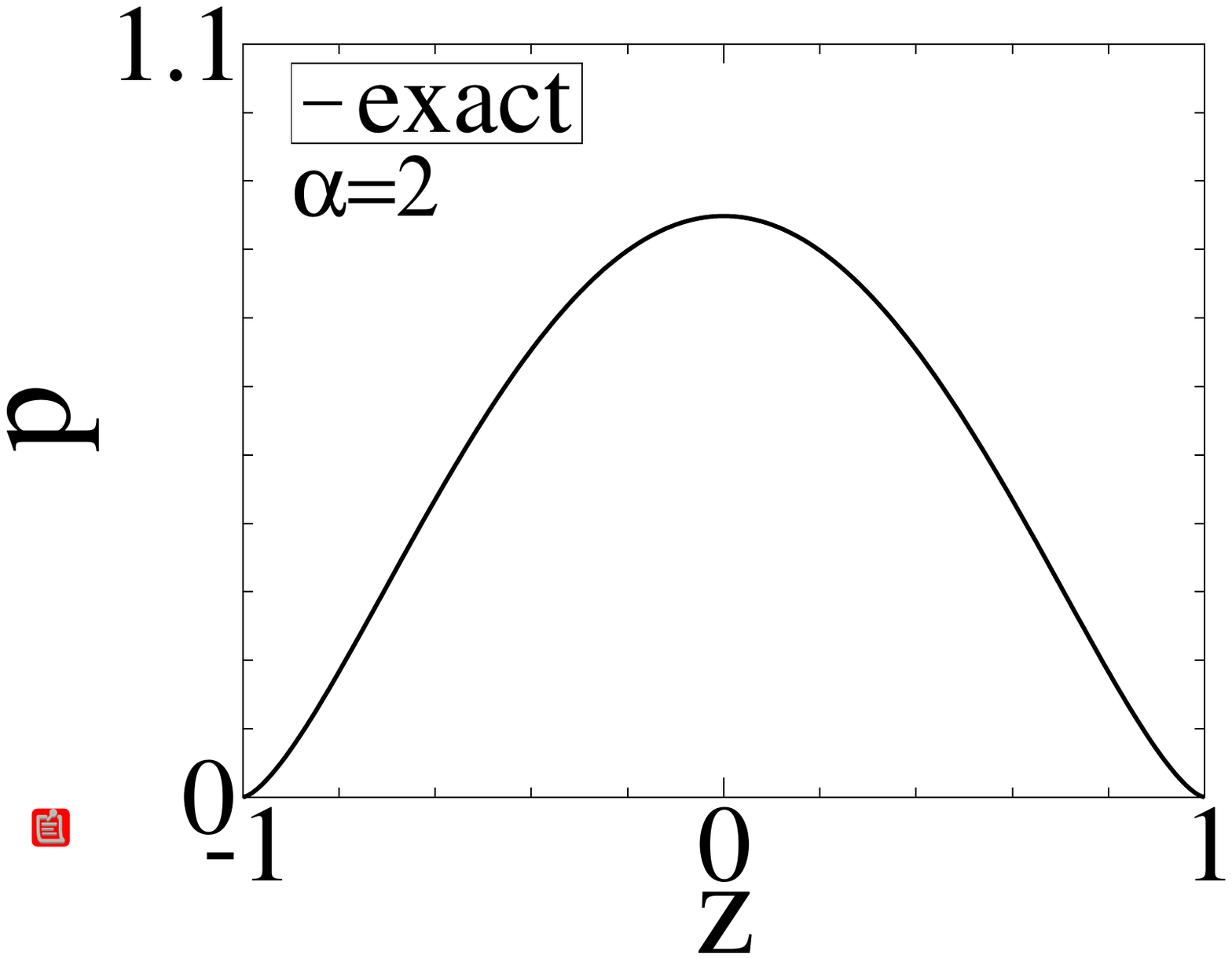} &
\hspace{-0.2cm}\includegraphics[height=0.16\textwidth,width=0.17\textwidth]{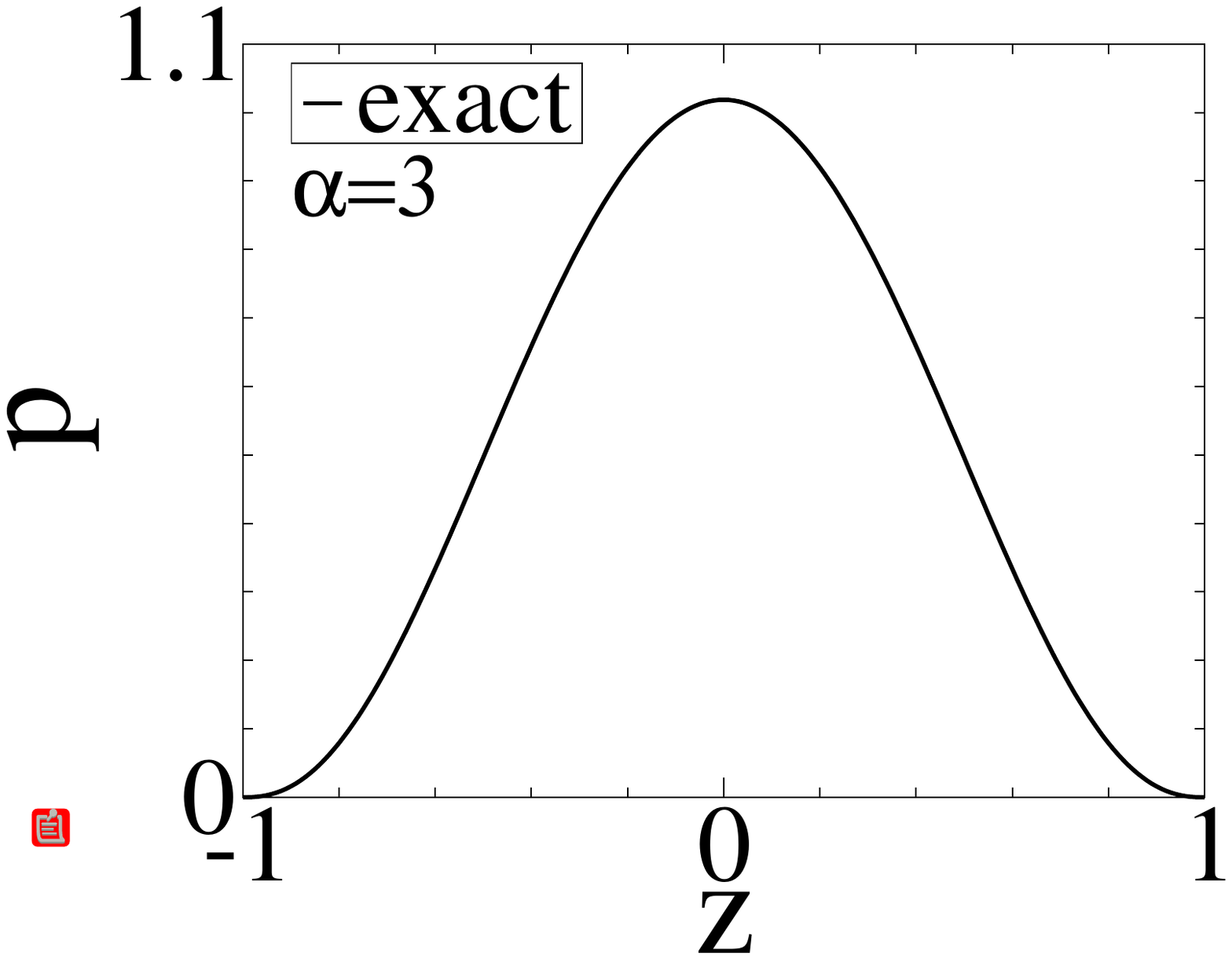} 
 \end{tabular}
 \end{center} 
\caption{ Distributions $p_w$ for different values of $\alpha$.    
Exact distributions (represented by lines) are obtained from Eq. (\ref{eq:pw-2dl}).  The circular symbols 
represent simulation data points.  In addition, below each plot for $p_w$ we plot a corresponding 
distribution $p$ to emphasize qualitatively different behavior.}
\label{fig:fig4} 
\end{figure}

In Fig. (\ref{fig:fig4}) we plot a number of different distributions $p_w$ for different values of $\alpha$ calculated
using Eq. (\ref{eq:pw-2dl}), where the integral is evaluated numerically.  Those distributions are compared with 
those obtained from simulations.  
{Simulations were carried out using the Euler method for updating particle positions:  
$$
x(t+\Delta t) = \left[ v_0 \cos\theta(t) - \mu K x(t)\right]\Delta t, 
$$
with new orientation $\theta(t)$ selected after a time period selected randomly from the Poisson 
distribution $P \propto e^{-t/\tau}$.   }
For comparison, in Fig. (\ref{fig:fig4}) we also plot the corresponding distributions $p$ below each $p_w$. 
\graphicspath{{figures/}}

Unlike the distributions in Fig. (\ref{fig:fig1}) for the system in 1D, a peak at $w=0$ does not fall to zero 
for $\alpha$ above the crossover.  We can calculate the height of $p_w$ at $w=0$ from Eq. (\ref{eq:pw2DL}) 
by setting $w$ to zero. This yields $p_w(0)  =  \frac{A}{2\pi} \frac{\alpha}{\alpha-1}  I(0)$ which simplifies to 
\be
p_w(0) =   
 \frac{1}{\alpha-1} \left[ \frac{4^\alpha}{\pi}   \frac{\alpha!\alpha!} {(2\alpha)!}\right]^2.  
\label{eq:pw0-2dl}
\ee
Rather than suddenly falling to zero for $\alpha>1$, the peak height at $w=0$ approaches zero 
algebraically as a function of $\alpha$.

For half integer values of $\alpha$, it is possible to obtain
exact expression for $p_w$.  Those expressions are derived in Appendix (\ref{sec:app1}).

Note that the crossover in 2D occurs at $\alpha=1$ while that in 1D
at $\alpha=2$.  Yet if we recall the discussion below Eq. (\ref{eq:alpha}), 
the actual rate for a system in 1D is not $\alpha$ but $\alpha/2$.   Therefore, 
considering the actual rates, the crossover in both dimensions occurs for the same 
rates.

\subsection{moment analysis}

The moments of the distribution $p(z)$ in Eq. (\ref{eq:pz-2dl}) are obtained directly from the formula
$\langle z^{2n}\rangle = \int_{-1}^1 dz\, p(z) z^{2n}$:  
\be
\langle z^{2n}\rangle = \frac{ \Gamma \left(n+\frac{1}{2}\right)  \Gamma (\alpha + 1)}  { \sqrt{\pi }  \Gamma ( \alpha + n + 1)}.  
\label{eq:z2n-2dl}
\ee
The moments generate a monotonically decreasing sequence whose infinite sum is 
\be
\sum_{n=0}^{\infty} \langle z^{2n}\rangle  
= \left\langle \frac{1}{1-z^2} \right\rangle 
= 
  \begin{cases}
      \frac{2\alpha}{2\alpha-1} & \text{if $\alpha>\frac{1}{2}$}\\
    \infty  & \text{if $\alpha\leq \frac{1}{2}$}.  
  \end{cases}
\label{eq:sum-2dl}
\ee
The sum diverges at $\alpha= \frac{1}{2}$.  This is linked to divergences in $p$ and does not represent a 
crossover.  An actual crossover is determined from the behavior of $p_w$.

\section{RTP particles in 3D:  linear harmonic trap}
\label{sec:sec3}

For RTP particles in a harmonic trap in 3D, theres is no available solution for a stationary distribution $p$.  
Instead of trying to obtain such a distribution directly, in this section we focus on how to obtain exact expressions fo 
the moments of $p$.  The results of this section are the most important results of this work.

\subsection{moment analysis}


The stationary Fokker-Planck equation for RTP particles in a harmonic potential $u=\frac{K}{2} x^2$
embedded in 3D is obtained from Eq. (\ref{eq:FP-RTP}) for the force ${\bf F} = -Kx{\bf e}_x$ and 
$$
\hat L\, \rho = \frac{1}{\tau} \left[ -\rho + \frac{1}{2} \int_0^{\pi} d\theta\, \sin\theta\, \rho(x,\theta) \right].  
$$
The resulting equation in reduced units is  
\be
0  =  \frac{\partial}{\partial z} \left[ \left( z  -  \cos\theta \right) \rho \right]  
-  \alpha\rho      +     \frac{\alpha}{2}   p, 
\label{eq:FP-LH3D}
\ee
with the marginal distribution defined as 
\be
p(z) =   \int_0^{\pi} d\theta \,  \sin\theta\, \rho(z,\theta).  
\label{eq:rho2pz-3d}
\ee


The Fokker-Planck equation in Eq. (\ref{eq:FP-LH3D}) can be transformed into the following recurrence relation: 
\be
A_{l,m}      =       \frac{\alpha }{ l+\alpha}   A_{l,0} A_{0,m}      +       \frac{l}{ l + \alpha } A_{l-1,m+1}
\label{eq:Alm2}
\ee
where 
$$
A_{l,m} = \langle z^l \cos^m\theta\rangle, 
$$
and the angular brackets indicate averaging procedure defined as 
$\langle \dots \rangle = \int_{-1}^{1} dz \int_0^{\pi} d\theta \, \rho \sin\theta (\dots)$.  

The recurrence relation reflects the structure of the differential equation from which it was obtained:    
it is first order, linear, with variable coefficients.  The relation was obtained by multiplying the 
Fokker-Planck equation by $z^l\cos^m\theta$ followed by integration 
$\int_{-1}^{1} dz\int_0^{\pi} d\theta \, \sin\theta$ and written in its final form 
using integration by parts.

Since $A_{l,0} = \langle z^{l}\rangle$, solving the recurrence relation would permit us to obtain moments.  
The initial condition of the recurrence relation is provided by the terms $A_{0,m}$ which are easily evaluated
$$
A_{0.m} =  \frac{1}{2} \int_0^{\pi} d\theta\, \sin \theta \cos^{m}\theta = 
  \begin{cases}
      \frac{1}{m+1} & \text{if $m$ even}\\
      0 & \text{if $m$ odd}.
  \end{cases}
$$

The recurrence relation cannot be solved for an arbitrary $A_{l,m}$.  Nonetheless, it is possible to 
reduce the relation to another recurrence relation in terms of $A_{2n,0}=\langle z^{2n} \rangle$ only:
\be
\langle z^{2n} \rangle  =  \frac{\alpha}{2n} \sum _{k=0}^{n-1}   \frac{ \langle z^{2k}\rangle}{2n-2k+1}  \frac{(2 k+1)_{\alpha-1}}{(2 n+1)_{\alpha-1}}.  
\label{eq:z2n-3dl}
\ee
where $(x)_n = \frac{\Gamma(x+n)} {\Gamma(x)}$ is the falling factorial.

The recurrence relation in Eq. (\ref{eq:z2n-3dl}) is the central result of this section.  
Although it does not provide an exact expression for an arbitrary moment, 
it provides an analytically tractable procedure for obtaining such an expression recursively.  
A number of initial even moments generated from the recurrence relation in Eq. (\ref{eq:z2n-3dl}) are given in Table (\ref{table2}).  
\begin{table}[h!]
\centering
 \begin{tabular}{ l l l } 
  \hline
 & \\[-1ex]
   ~~~  {$ \langle z^2\rangle = \frac{1}{3} \frac{1}{1+\alpha}$ }   \\ [1.ex] 
   ~~~  {$ \langle z^4\rangle = \frac{1}{15} \frac{18 + 5 \alpha}{ (1 + \alpha) (2 + \alpha ) (3 + \alpha)}$}    \\ [1.ex] 
   ~~~  { $ \langle z^6\rangle = \frac{1}{63} \frac{ 1080 + 378 \alpha + 35 \alpha ^2}{(\alpha +1) (\alpha +2) (\alpha +3) (\alpha +4) (\alpha +5)} $}   \\ [1.ex] 
   ~~~  { $ \langle z^8\rangle = \frac{1}{135} \frac{ 75600 + 28404 \alpha   + 3780 \alpha^2 + 175 \alpha^3 }{(\alpha +1) (\alpha +2) (\alpha +3) (\alpha +4) (\alpha +5) (\alpha + 6) (\alpha + 7) } $}   \\ [1.ex] 
   ~~~  { $ \langle z^{10}\rangle = \frac{1}{99} \frac{ 3265920 + 1259280 \alpha   +  193644  \alpha^2 +  13860 \alpha^3  +  385 \alpha^4} 
   {(\alpha +1) (\alpha +2) (\alpha +3) (\alpha +4) (\alpha +5) (\alpha + 6) (\alpha + 7) (\alpha + 8) (\alpha + 9) } $}   \\ [1.ex] 
  \hline
\end{tabular}
\caption{Moments of a stationary distribution $p$ obtained from a recurrence relation in Eq. (\ref{eq:z2n-3dl}).  }
\label{table2}
\end{table}

By examining Table (\ref{table2}), we can verify that  for $\alpha=0$ the moments reduce to a simple general formula 
$\langle z^{2n}\rangle = \frac{1}{2n+1}$ that can be linked to a uniform distribution where $\langle z^{2n}\rangle = \frac{1}{2} \int_{-1}^{1}dz\, z^{2n}$.  
This means that for finite $\alpha$, $p$ can only be convex which, in turn, implies the absence of divergences at the trap borders.  

To understand how a flat distribution arises for $\alpha=0$
we should understand that for $\alpha=0$ all particles are immobile and trapped at $z=\cos\theta$, where the swimming 
velocity and the velocity due to harmonic potential 
cancel one another.  As a result $\rho = \frac{1}{2} \delta(z-\cos\theta)$.  Averaged 
over all orientations, this yields:  
\be
\lim_{\alpha\to 0} p(z)   = \frac{1}{2} \int_0^{\pi} d\theta \, \sin\theta \, \delta(z-\cos\theta) = \frac{1}{2}.  
\label{eq:pz-a0}
\ee
The averaging procedure completely smooths out the delta distribution.

{In the Table (\ref{table2a}), we compare the second and fourth moments calculated for different dimension.  
For 1D and 2D, these moments are obtained from the formulas in Eq. (\ref{eq:z2n-1d}) and Eq. (\ref{eq:z2n-2dl}). 
The tendency is that an increased dimensionality reduces the value of a given moment.  The best way to understand 
this reduction is to think of each system as one-dimensional with different distributions of swimming velocities
(for a potential $u = \frac{K}{2} x^2$ we actually consider the projection of a swimming velocity along the $x$-axis
and not the true swimming velocity).  
\begin{table}[h!]
\centering
 \begin{tabular}{ l l l } 
  \hline
 & \\[-2ex]
 &  ~~~~~ $\langle z^2\rangle$ &  $~~~~~~~~~ \langle z^4\rangle$   \\ [0.2ex] 
 & \\[-2.2ex]
  \hline
 & \\[-1ex]
1D  & ~~~   {~\,\,$\frac{1}{1+\alpha}$ }   &  {~~\,$ \frac{3}{(1+\alpha)(3+\alpha)}$ }   \\ [1.ex] 
2D  & ~~~   {$\frac{1}{2} \frac{1}{1+\alpha} $}  &   {~$\frac{1}{4} \frac{3}{(1+\alpha)(2+\alpha)}$ }    \\ [1.ex] 
3D  & ~~~   {$\frac{1}{3} \frac{1}{1+\alpha} $}  &   {~$\frac{1}{15} \frac{18 + 5 \alpha}{ (1 + \alpha) (2 + \alpha ) (3 + \alpha)}$ }   \\ [1.ex] 
  \hline
\end{tabular}
\caption{Second and fourth moments of the distribution $p$ in different dimensions.  }
\label{table2a}
\end{table}
}

\subsection{distribution in $w$-space}

To obtain a distribution in $w$-space we follow a similar procedure to that used in Sec. (\ref{sec:sec2A}).  
We transform Eq. (\ref{eq:FP-LH3D}) using the change of variables 
$z=-w+\cos\theta$.  The resulting equation is then integrated over all orientations.  The procedure yields the first order 
inhomogeneous differential equation:   
\be
0 =      (1 -  \alpha) p_w    +   w p_w'   +    \frac{\alpha}{2} \int_{-1}^{1-w} ds \,  p(s), 
\label{eq:diff3DL}
\ee
for which the solution is 
\be
p_w   =   \frac{\alpha}{2}  |w|^{\alpha-1}  \left[ \int_{|w|}^{2} dw'\,  w'^{-\alpha} \int_{-1}^{1-w'} ds \,  p(s) \right].  
\label{eq:pw3DL}
\ee
The solution permits us to obtain $p_w$ from $p$.  The difference between this result and that in 
Eq. (\ref{eq:pw-2dl}) is that here we do not have the exact expression for $p$.  

Even without knowing $p$, Eq. (\ref{eq:diff3DL}) can be used to calculate  
$p_w(0)$ by setting $w=0$.  This yields 
\be
p_w(0)    =    \frac{1}{2}  \frac{\alpha}{\alpha-1}.  
\label{eq:pw0-3dl}
\ee
The expression diverges at $\alpha=1$, indicating a divergence in $p_w$ and the presence of nearly 
immobile particles.  Compared with a similar result in Eq. (\ref{eq:pw0-2dl}) for a system in 2D, we 
see that the divergence occurs at the same value of $\alpha$.   This means that the location of the crossover
is independent of the system dimension. 

\begin{figure}[hhhh] 
 \begin{center}
 \begin{tabular}{rrrr}
\hspace{-0.3cm}\includegraphics[height=0.15\textwidth,width=0.16\textwidth]{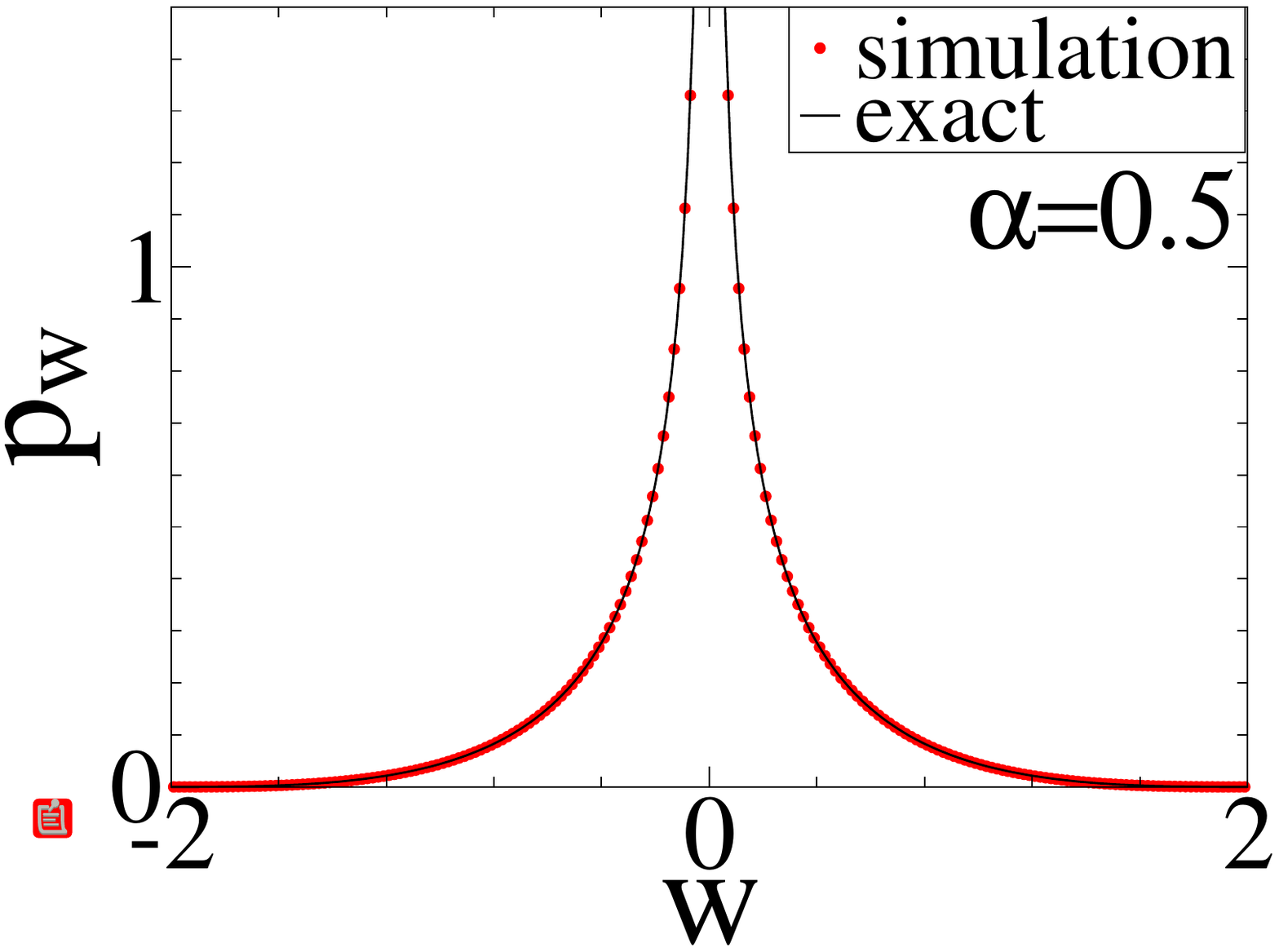} &
\hspace{-0.2cm}\includegraphics[height=0.15\textwidth,width=0.16\textwidth]{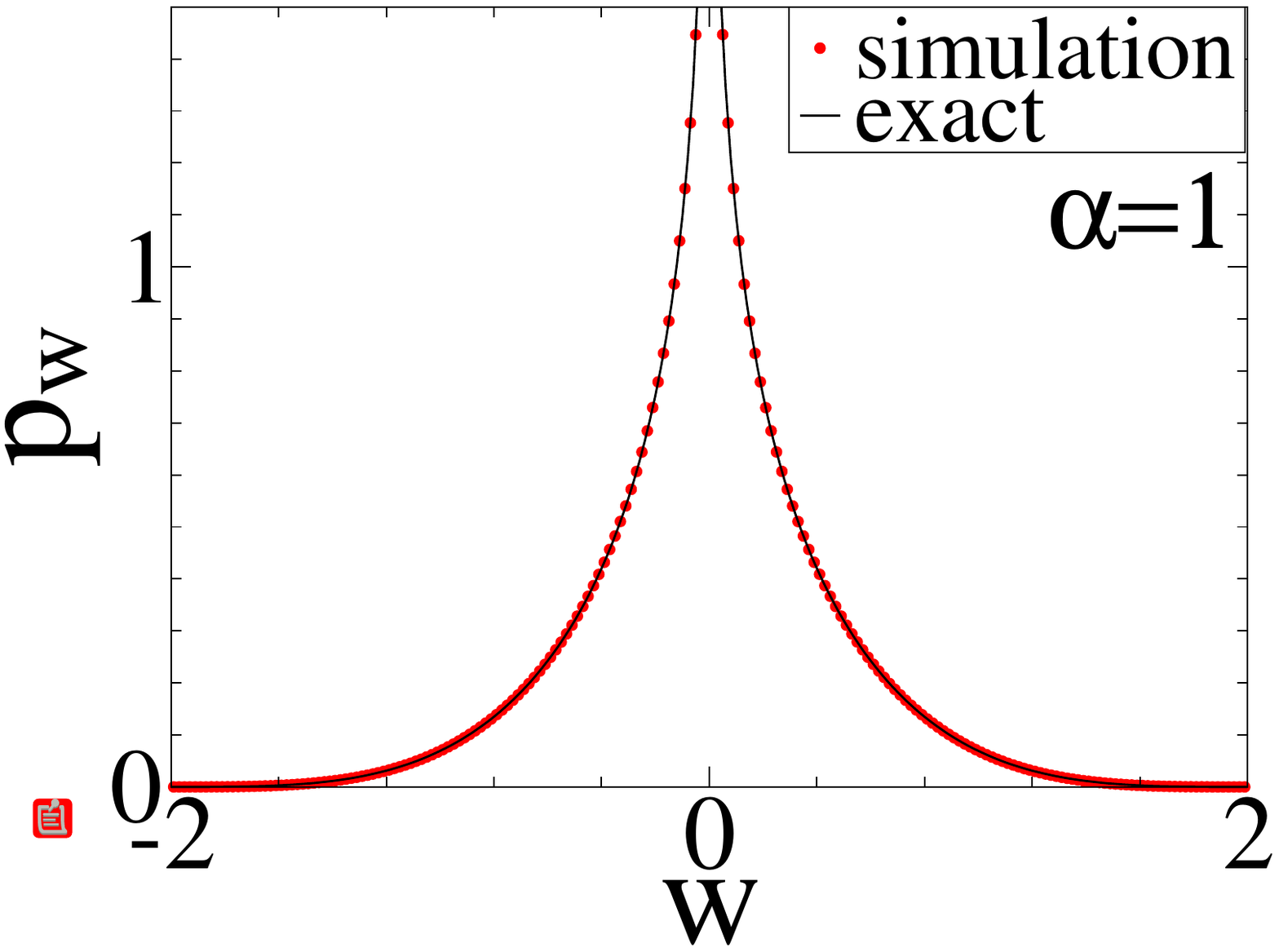} &
\hspace{-0.2cm}\includegraphics[height=0.15\textwidth,width=0.16\textwidth]{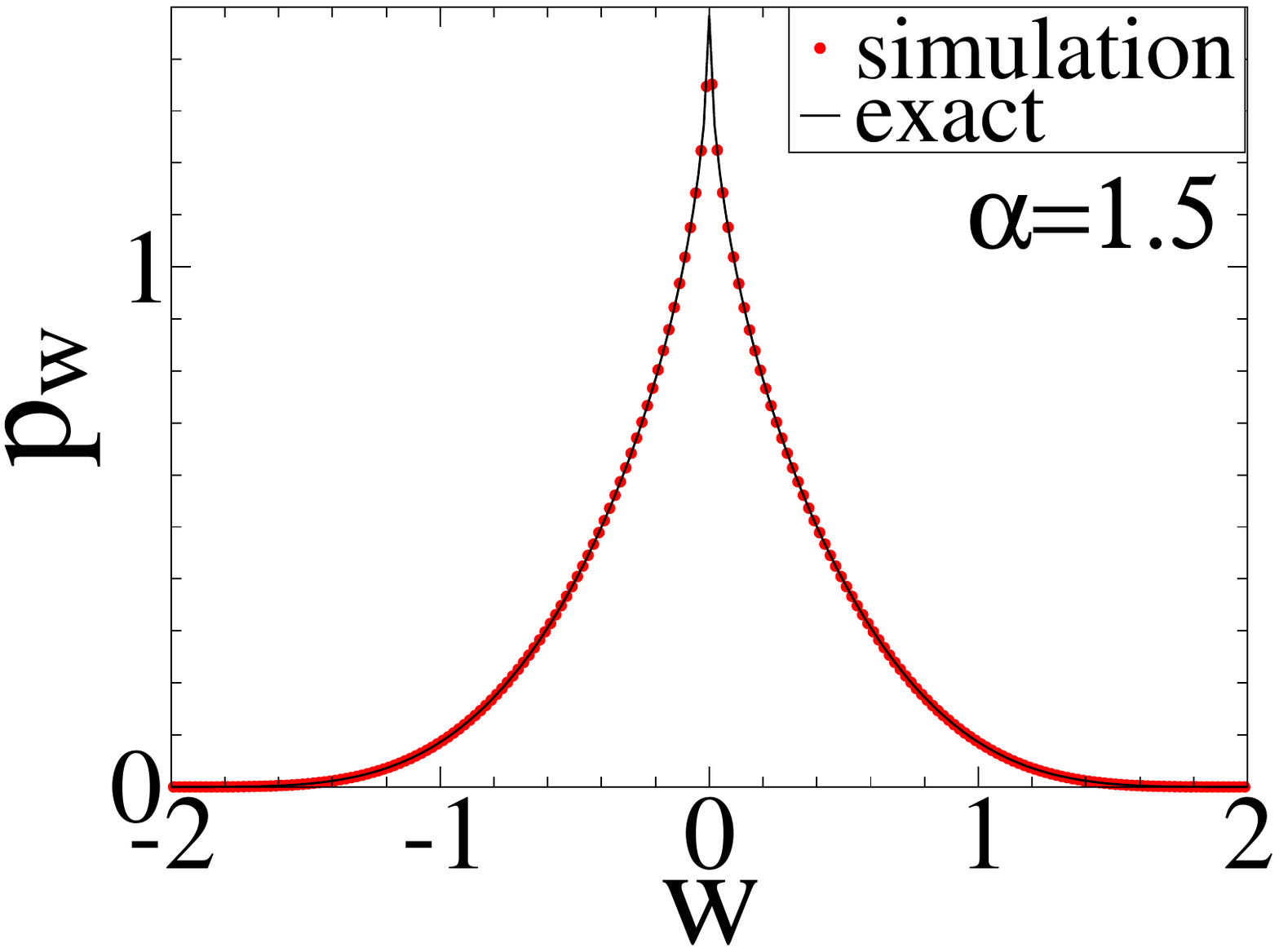} \\
\hspace{-0.3cm}\includegraphics[height=0.15\textwidth,width=0.16\textwidth]{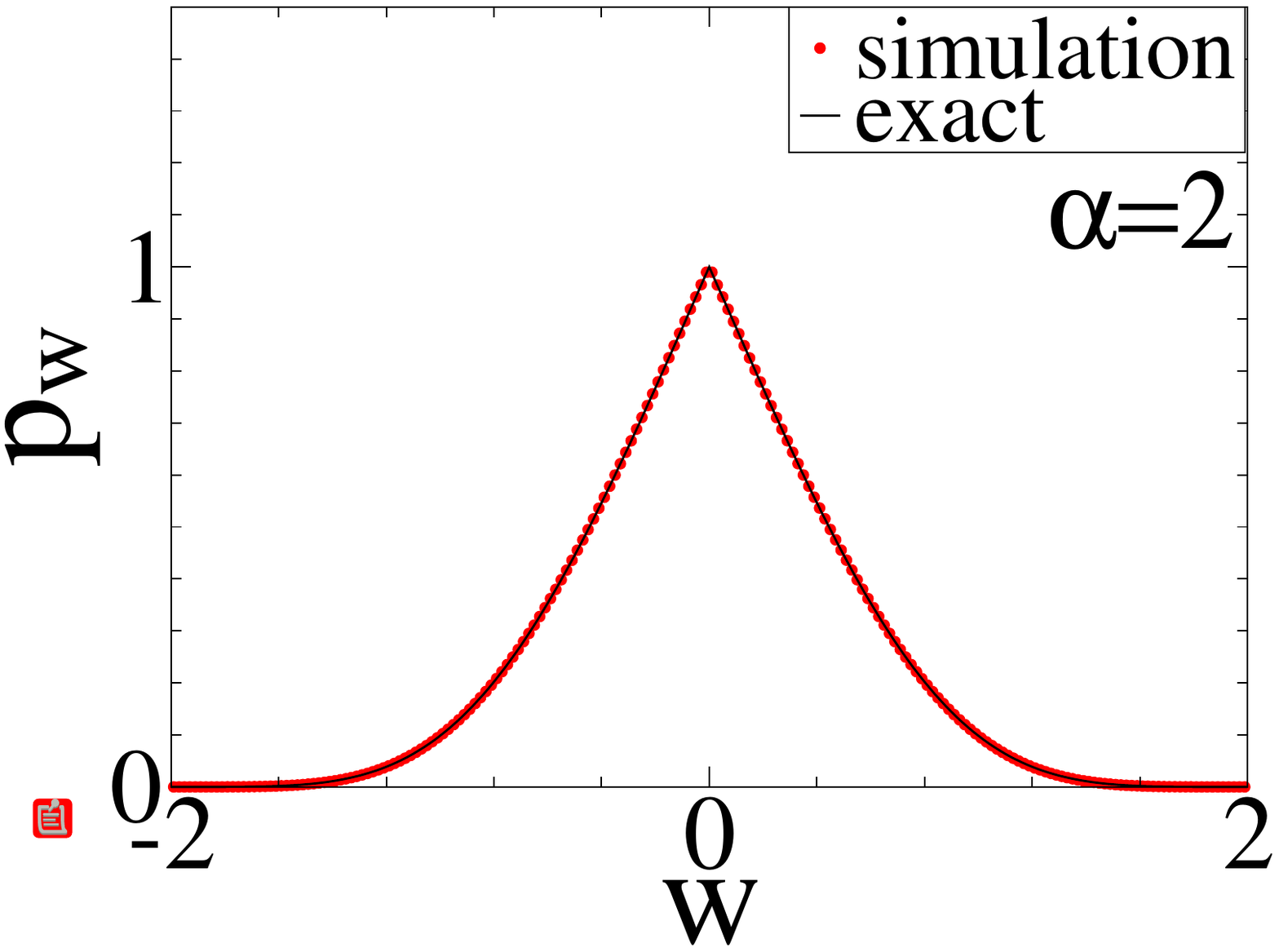} &
\hspace{-0.2cm}\includegraphics[height=0.15\textwidth,width=0.16\textwidth]{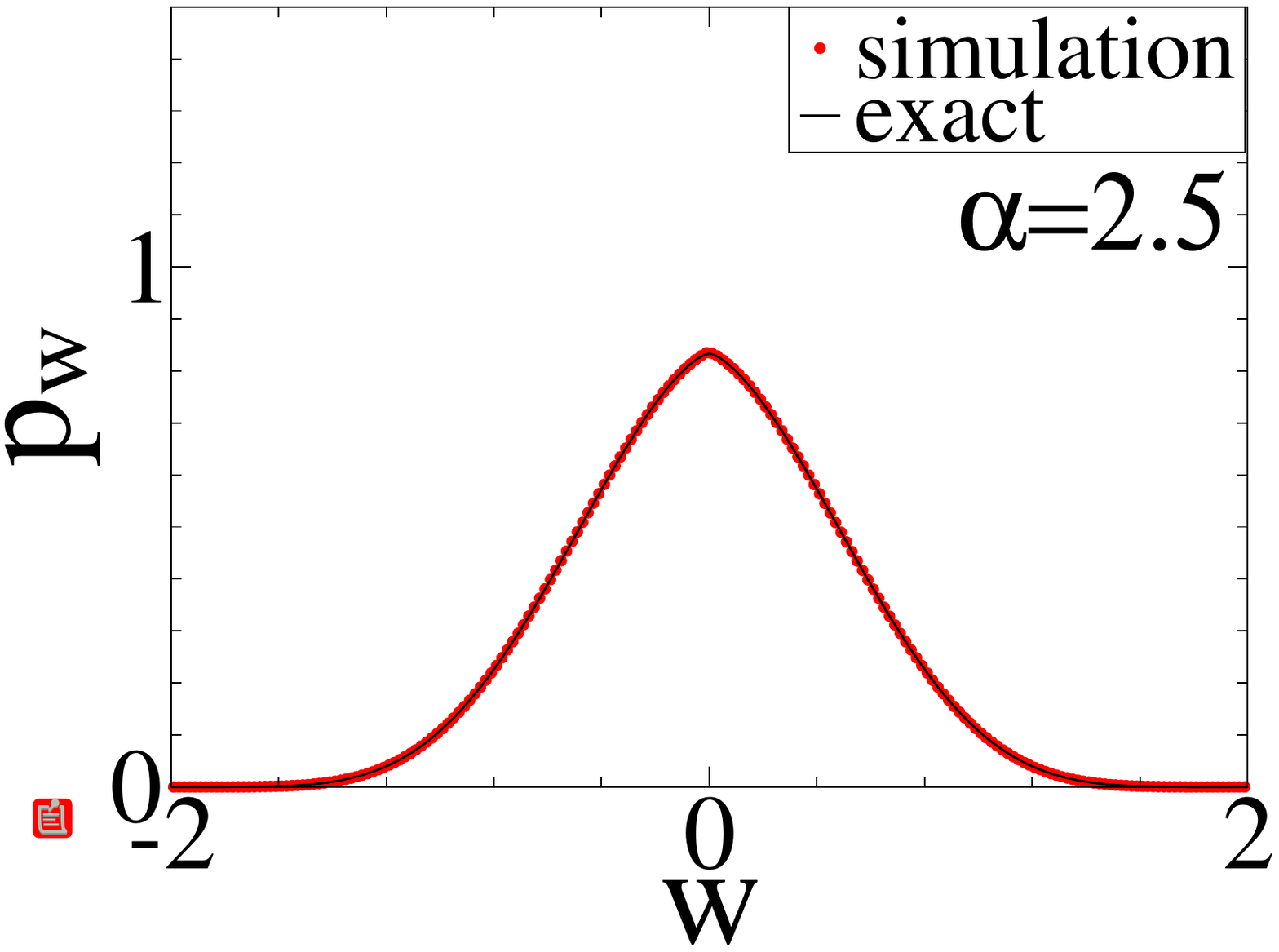} &
\hspace{-0.2cm}\includegraphics[height=0.15\textwidth,width=0.16\textwidth]{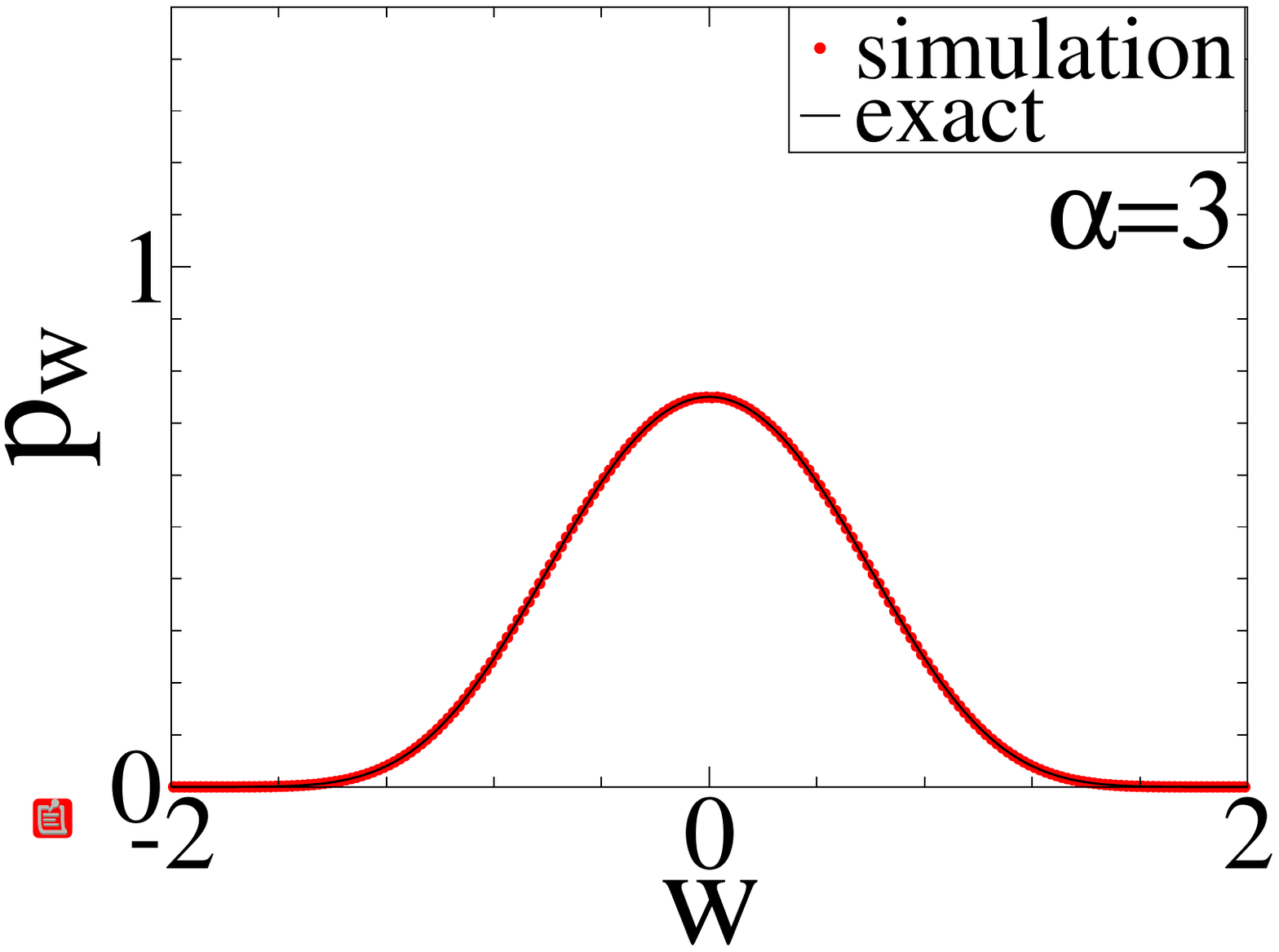} 
 \end{tabular}
 \end{center} 
\caption{Distributions $p_w$ calculated from $p$ using numerically evaluated Eq. (\ref{eq:pw3DL}).  
The height of the peak at $w=0$ for $\alpha>1$ is given by Eq. (\ref{eq:pw0-2dl}).} 
\label{fig:fig9} 
\end{figure}

In Fig. (\ref{fig:fig9}) we plot $p_w$ for different values of $\alpha$ obtained using Eq. (\ref{eq:pw3DL}).  
The integrals are calculated numerically and $p$ is calculated from the moments as explained in the 
next section.  
\graphicspath{{figures/}}

\subsection{recovering $p$ from moments}
\label{sec:rec-p3dl}

The recurrence relation in Eq. (\ref{eq:z2n-3dl}) permits fast computation of an arbitrary number 
of even moments of $p$.  In this section we present a procedure for recovering a distribution $p$ 
from the moments based on the Fourier-Legendre expansion, 
\be
p(z) = \sum_{n=0}^{\infty} a_n P_{2n} (z), 
\label{eq:pz-Pn}
\ee
where $P_m$ are Legendre (orthogonal) polynomials.  
Like $p$, Legendre polynomials are defined on $[-1,1]$.  
Due to even symmetry of $p$, only even Legendre polynomials $P_{2n}$ are required: 
\be
P_{2n} = 2^{2 n} \sum _{k=0}^n \frac{ z^{2 k} \,  \Gamma \left(k+n+\frac{1}{2}\right) } { (2 k)! (2 n-2 k)! \,  \Gamma \left(k-n+\frac{1}{2}\right)}.  
\label{eq:P2n}
\ee
The coefficients $a_n$ in Eq. (\ref{eq:pz-Pn}) can be determined from the orthogonality relation $ \int_{-1}^{1} dz\, P_{n}P_m = \frac{2}{2n+1} \delta_{mn}$
which leads to 
\be
a_n = \frac{ 4n + 1 }{2}  \int_{-1}^{1} dz\, P_{2n} (z) p(z), 
\ee
and in combination with Eq. (\ref{eq:P2n}) yields
\be
a_n = \frac{4 n+1}{2}  \left[ 2^{2 n} \sum _{k=0}^n \frac{  \langle z^{2 k} \rangle \,  \Gamma \left(k+n+\frac{1}{2}\right) } { (2 k)! (2 n-2 k)! \,  \Gamma \left(k-n+\frac{1}{2}\right)} \right].  
\label{eq:an}
\ee

The expansion in Eq. (\ref{eq:pz-Pn}) together with the coefficients in Eq. (\ref{eq:an}) provides an 
exact formula for recovering $p$ in terms of moments obtained from Eq. (\ref{eq:z2n-3dl}).
Initial coefficients $a_n$ are listed in Table (\ref{table3}).  
\begin{table}[h!]
\centering
 \begin{tabular}{ l l l } 
  \hline
 & \\[-1ex]
   ~~~  { $ a_0 = \frac{1}{2}$ }   \\ [1.ex] 
   ~~~  { $ a_1 =  -\frac{5}{4} \frac{\alpha}{ \alpha + 1} $}    \\ [1.ex] 
   ~~~  { $ a_2 =  -\frac{3}{16}  \frac{16 \alpha  - 24 \alpha^2 - 9 \alpha^3}{ (\alpha + 1)(\alpha + 2) (\alpha + 3) } $}   \\ [1.ex] 
   ~~~  { $ a_3 =  -\frac{13}{96}  \frac{ 288 - 496 \alpha + 120 \alpha^2 + 120 \alpha^3 + 15 \alpha^4 }{ (\alpha + 1) (\alpha + 2) (\alpha + 3)  (\alpha + 4) (\alpha + 5) } $}   \\ [1.ex] 
   ~~~  { $ a_4 =  -\frac{17}{768}   \frac{ 55296 - 105984 \alpha + 49280 \alpha^2 + 8512 \alpha^3 - 6720 \alpha^4  - 1680 \alpha^5  - 105 \alpha^6} 
   {(\alpha +1) (\alpha +2) (\alpha +3) (\alpha +4) (\alpha +5) (\alpha + 6) (\alpha + 7)  } $}   \\ [1.ex] 
  \hline
\end{tabular}
\caption{Coefficients for the Fourier-Legendre series in in Eq. (\ref{eq:pz-Pn}).  }
\label{table3}
\end{table}
Note that by setting $\alpha=0$, $a_0=1/2$ and all the remaining coefficients $a_n$ are zero.  
This implies a uniform distribution in agreement with the results in Eq. (\ref{eq:pz-a0}) for the same limit.

\begin{figure}[hhhh] 
 \begin{center}
 \begin{tabular}{rrrr}
\hspace{-0.3cm}\includegraphics[height=0.15\textwidth,width=0.16\textwidth]{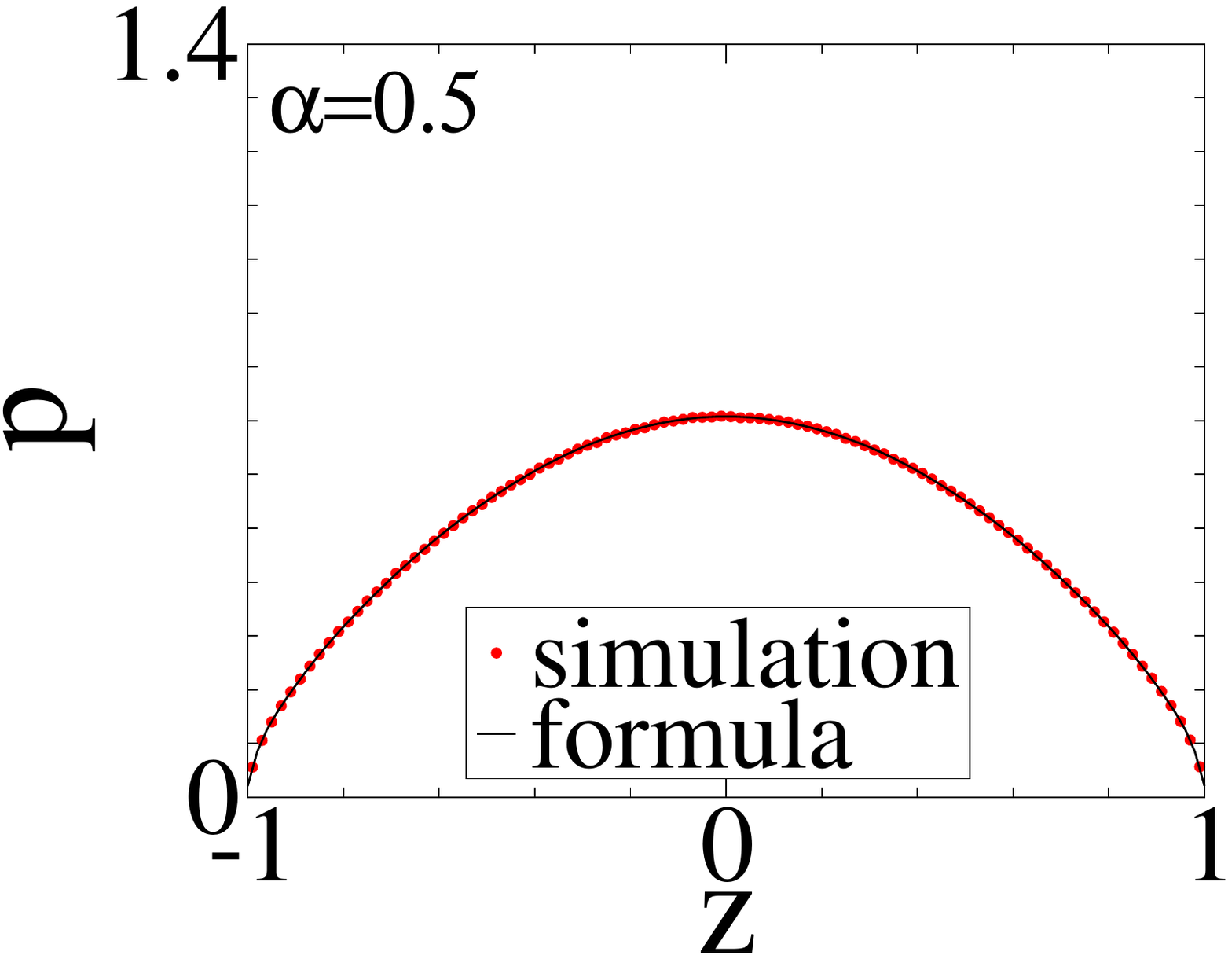} &
\hspace{-0.2cm}\includegraphics[height=0.15\textwidth,width=0.16\textwidth]{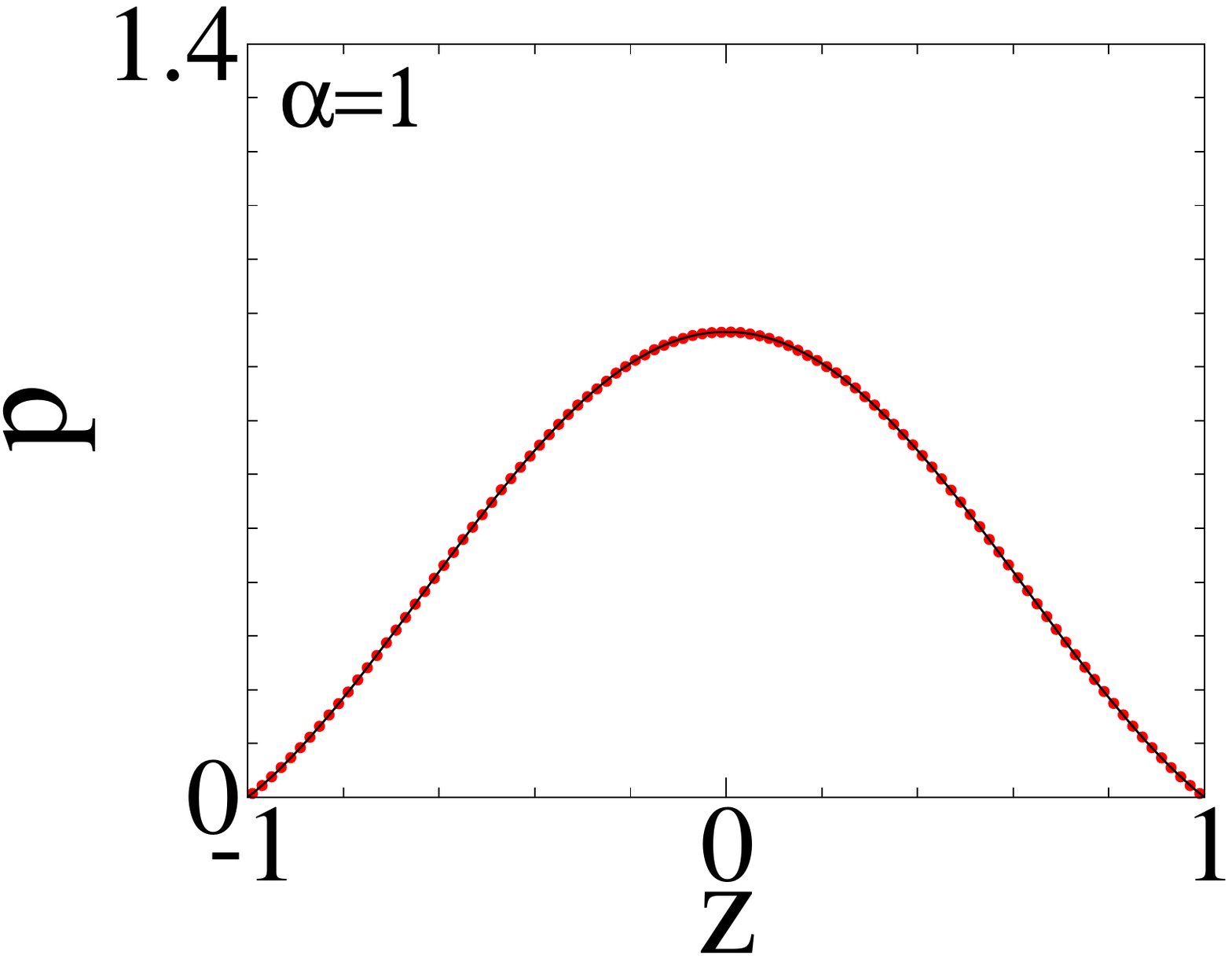} &
\hspace{-0.2cm}\includegraphics[height=0.15\textwidth,width=0.16\textwidth]{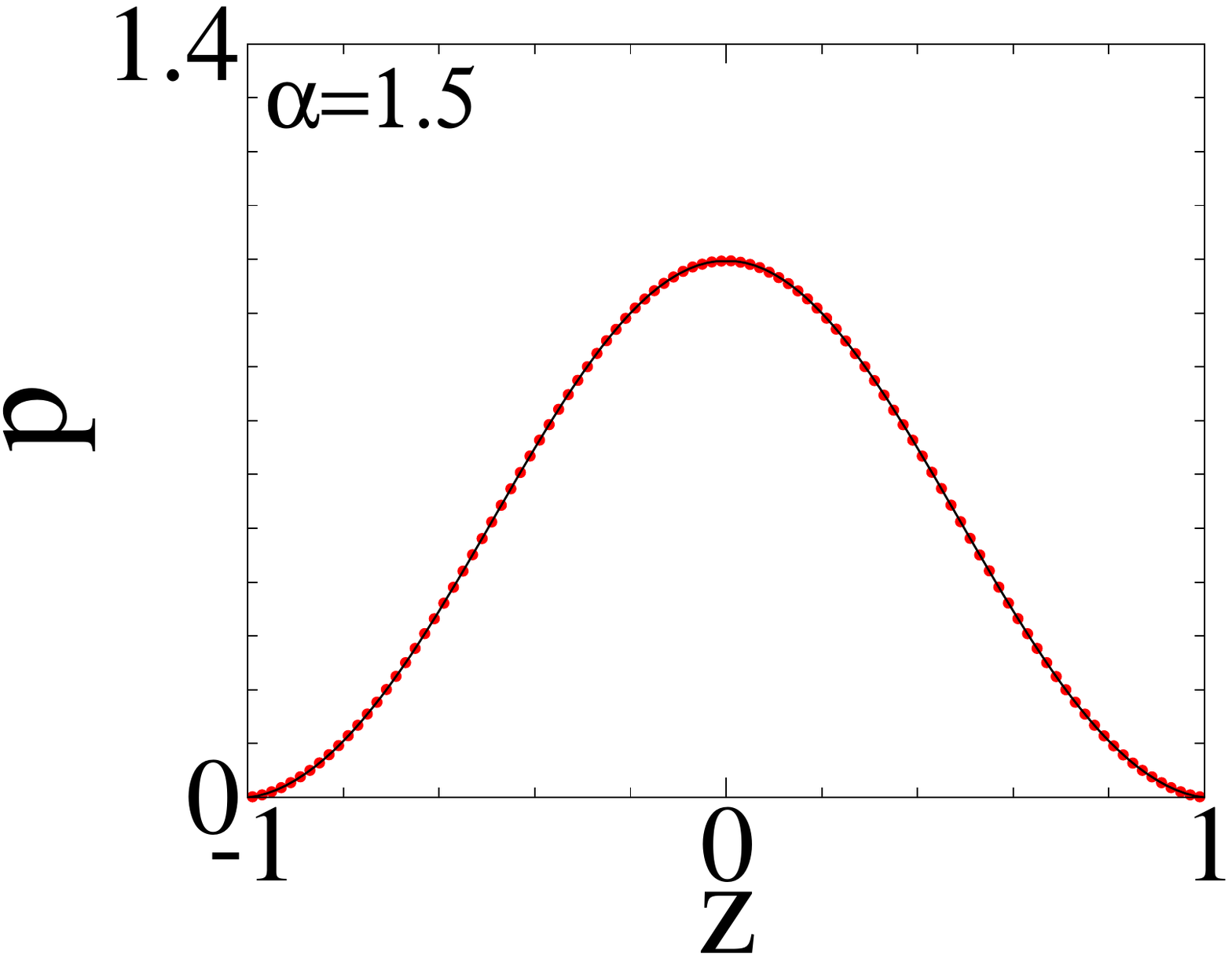} \\
\hspace{-0.3cm}\includegraphics[height=0.15\textwidth,width=0.16\textwidth]{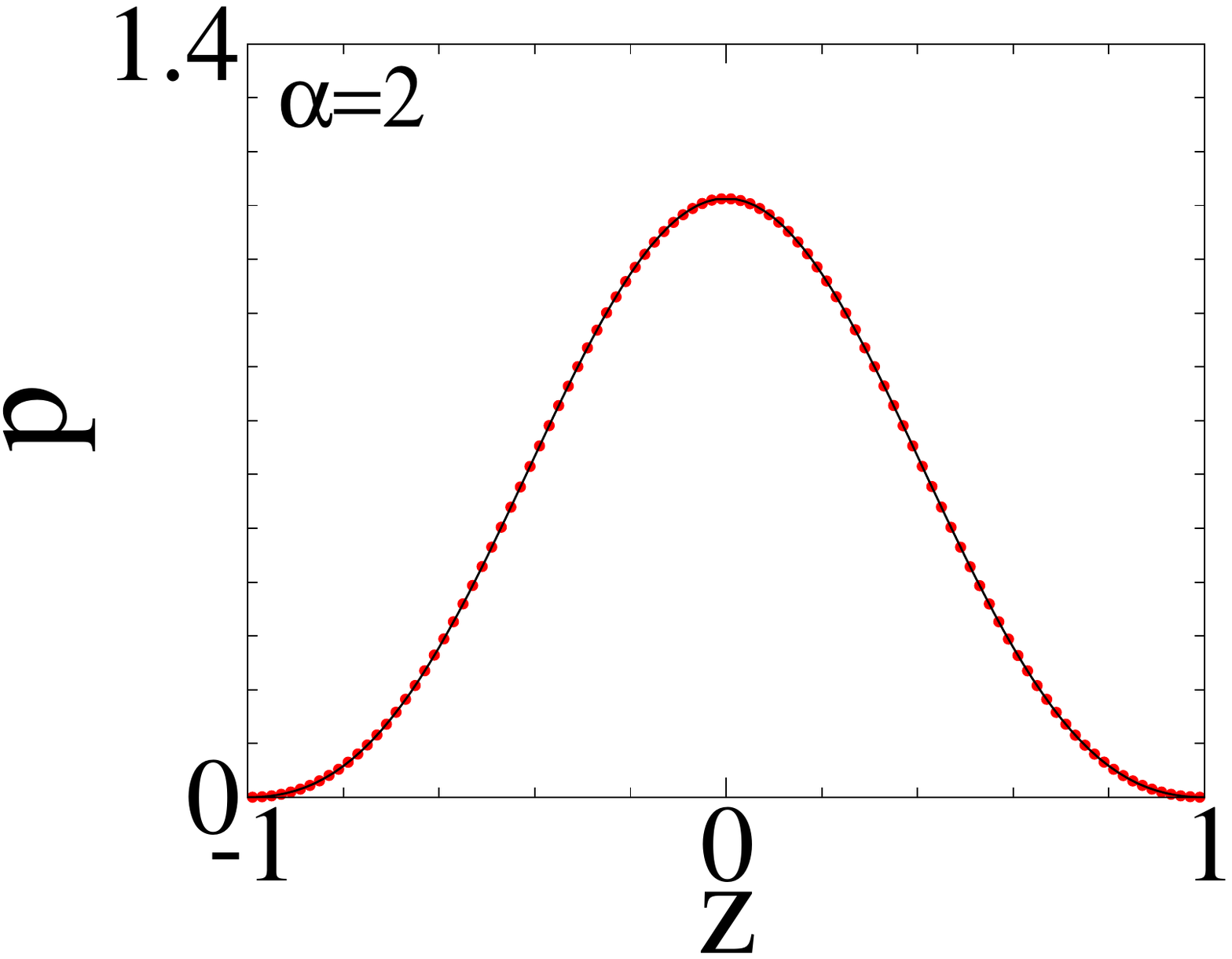} &
\hspace{-0.2cm}\includegraphics[height=0.15\textwidth,width=0.16\textwidth]{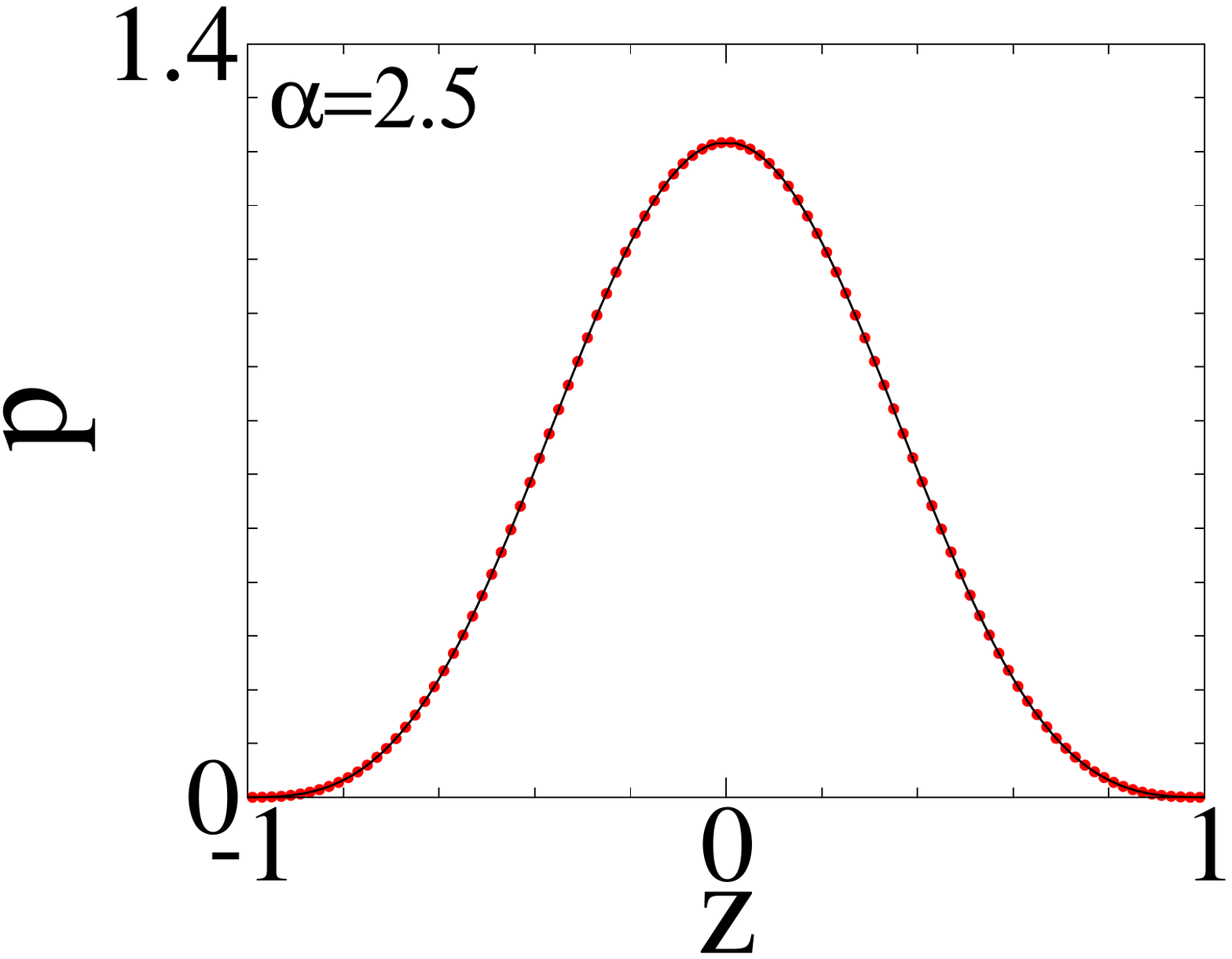} &
\hspace{-0.2cm}\includegraphics[height=0.15\textwidth,width=0.16\textwidth]{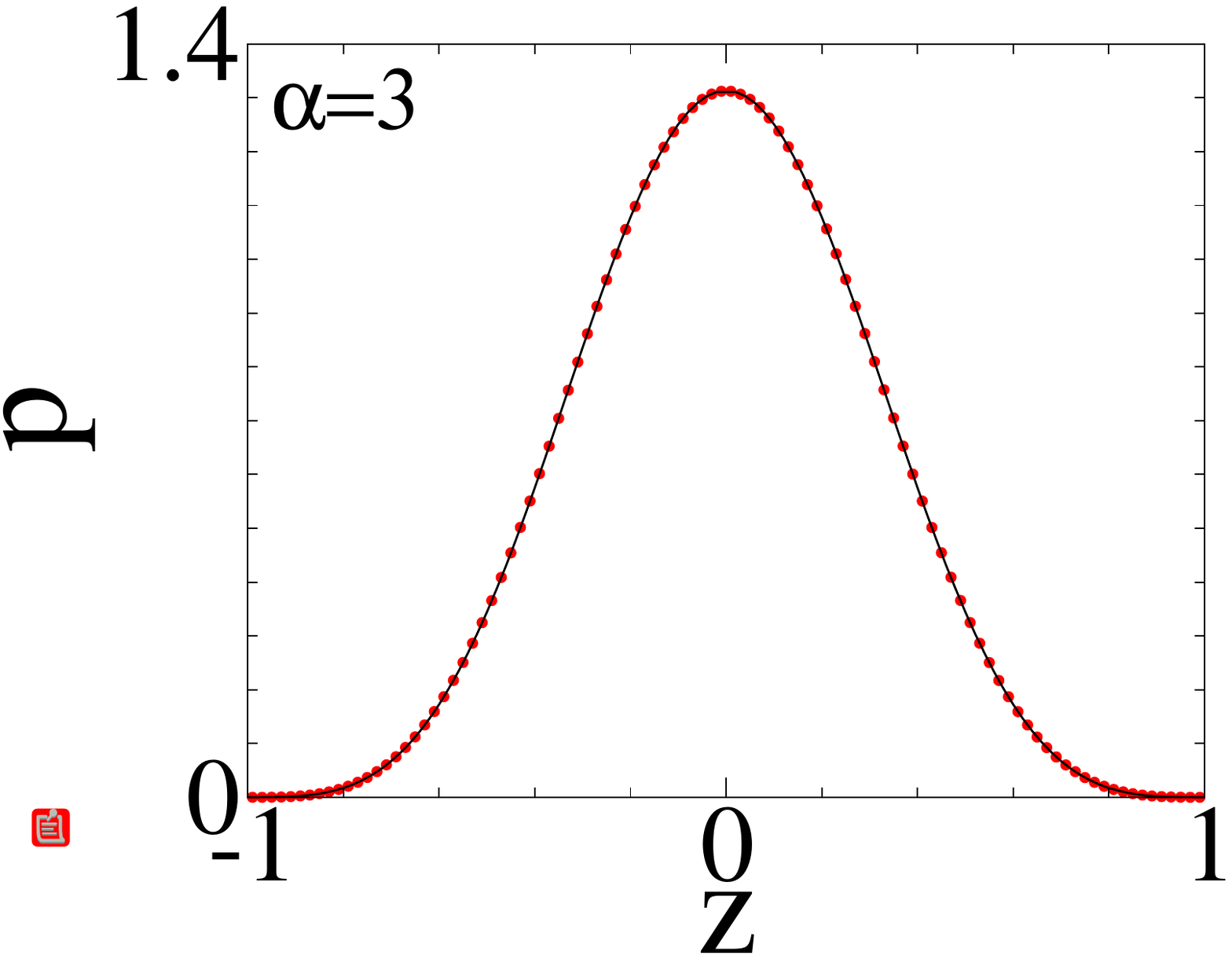} 
 \end{tabular}
 \end{center} 
\caption{Distributions $p$ calculated using the truncated Fourier-Legendre series $p =  \sum_{n=0}^{N_c} a_n P_{2n} (z)$, 
for $N_c=10$ and $a_n$ defined in Eq. (\ref{eq:an}).  }
\label{fig:fig8} 
\end{figure}

Recovered distributions obtained from a truncated Fourier-Legendre series $p =  \sum_{n=0}^{N_c} a_n P_{2n} (z)$
for $N_c=10$ are shown in Fig. (\ref{fig:fig8}).  
\graphicspath{{figures/}}

The truncated series shows a very good agreement with $p$ obtained from simulations.  
The larger the $\alpha$, the less terms of expansion are needed.  
It is generally recognized that the delta and square distributions are not well approximated by the 
series.   But since only the distribution at $\alpha=0$ is flat (and does not need to be approximated),
this limitation does not apply to our situation.

\section{Thermal fluctuations} 
\label{sec:sec-TF1D}

Up to this point, all the analysis and results were done at zero temperature.  To incorporate thermal fluctuations, 
we use a known result that for particles in a harmonic trap a stationary distribution of active particles in a thermal
bath can be represented as convolution \cite{Frydel22c}, 
for a system with 1D geometry given by 
\be
p_T(x) =  \int  d{x}'\, p(x') p_{eq}({x}-{x}'),
\label{eq:pT}
\ee
where 
$
p_{eq}({x}) =  \sqrt{ \frac{\mu K}{2\pi D} } e^{-\mu K x^2/2D} 
$
is the Boltzmann distribution of passive particles in a harmonic trap, and $p$ is the stationary distribution 
at zero temperature.  

A convolution construction applies to any combination of independent random processes 
\cite{book08,book19}.  Generally, however, confinement leads to correlations between random 
processes as it gives rise to nonlinear terms in the Langevin equation, 
even if these processes are originally independent.  The exception is a harmonic potential
whose force is linear and it does not introduce nonlinear terms.  
See Appendix (\ref{sec:app2}) for further discussion regarding this point.

Using Eq. (\ref{eq:pT}), the moments of the distribution $p_T$ are defined as 
\be
\langle z^{2n} \rangle_T =   \int_{-\infty}^{\infty} dz\, z^{2n} \int_{-1}^{1} dz' \, p(z') p_{eq}(z-z'),
\label{eq:z2nT0}
\ee
assuming dimensionless units.  
Then using the identity $z^{2n} = \left[ (z-z') + z' \right]^{2n}$ together with binomial expansion, 
Eq. (\ref{eq:z2nT0}) yields 
\be
\langle z^{2n} \rangle_T    =    \sum _{k=0}^{n} \frac{ (2 n)!  }{(2 k)! (2 n-2 k)!} \langle z^{2 n-2 k}\rangle  \langle z^{2 k} \rangle_{eq}. 
\label{eq:z2nTA}
\ee
And since moments $\langle z^{2 k} \rangle_{eq}$ can be calculated using the Boltzmann distribution, 
we finally get 
\be
\langle z^{2n} \rangle_T    =    \sum _{k=0}^{n} \frac{  (2 n)!  B^k }{ 2^{k}  k! (2 n-2 k)!} \langle z^{2 n-2 k}\rangle, 
\label{eq:z2nT}
\ee
where 
$
B =  \frac {\mu K} {v_0^2 }  D
$
is the dimensionless diffusion constant.  

Note that the moments for a finite temperature are given as an expansion in terms of moments at zero temperature.    
Since all terms in the expansion are positive, the effect of temperature is to increase the value of all the moments.  

{ Using Eq. (\ref{eq:z2nTA}), the initial moments are given by
\ba
&& \langle z^{2} \rangle_T    =    \langle z^{2}\rangle  +  \langle z^{2} \rangle_{eq} \nonumber\\
&& \langle z^{4} \rangle_T    =    \langle z^{4}\rangle   +   6 \langle z^{2}\rangle  \langle z^2 \rangle_{eq}  +   \langle z^{4} \rangle_{eq} \nonumber\\
&& \langle z^{6} \rangle_T    =    \langle z^{6}\rangle   
                                                    +   15 \langle z^{4}\rangle  \langle z^2 \rangle_{eq}    +   15 \langle z^{2}\rangle  \langle z^4 \rangle_{eq}  
+   \langle z^{6} \rangle_{eq} \nonumber\\
\ea
Note that the two contributions of the second moment are completely additive.  
Using Eq. (\ref{eq:z2nT}), the initial moments in the actual units are given by 
%
\ba
&& \langle x^{2} \rangle_T    =     \langle x^{2}\rangle  +   \frac{k_BT}{K} \nonumber\\
&&  \langle x^{4} \rangle_T    =     \langle x^{4}\rangle   +   6  \langle x^{2}\rangle  \frac{k_BT}{K}   +   3  \left(\frac{k_BT}{K}\right)^2 \nonumber\\
&& \langle x^{6} \rangle_T    =    \langle x^{6}\rangle   
                                                    +   15  \langle x^{4}\rangle   \frac{k_BT}{K}   +   45  \langle z^{2}\rangle  \left(\frac{k_BT}{K}\right)^2
+   15 \left(\frac{k_BT}{K}\right)^3 \nonumber\\
\ea
where we used $D= \mu k_BT$.  This result shows more clearly contribution of the temperature.  
}

\section{Isotropic harmonic trap}
\label{sec:sec4}

The previous analysis was done for a harmonic potential in a single direction, $u= \frac{1}{2} Kx^2$,
and it is not clear how and if the obtained results apply to an isotropic potential $u= \frac{1}{2} Kr^2$.  
In this section we extend the previous results to such an isotropic potential.

To establish a relation between moments for a linear potential, $\langle x^{2n}\rangle$, and the moments
of an isotropic potential, $\langle r^{2n}\rangle$, we consider first the Boltzmann distribution, 
$
p_{eq}(r) \propto e^{-\frac{\mu K r^2}{2 D}} 
$
to see how respective moments are related in this case.  For an arbitrary dimension $d$, the moments are 
easily evaluated and are given by 
\be
\langle r^{2n} \rangle   =   \left( \frac{2D} {\mu K} \right)^{n}   \frac{\Gamma\left( \frac{d}{2} + n \right) } {\Gamma\left( \frac{d}{2} \right) }.  ~~~~
\label{eq:r2n}
\ee
Note that $\langle x^{2n}\rangle = \langle r^{2n}\rangle_{d=1}$ so that we can write 
$$
\langle x^{2n}\rangle =  \left( \frac{2D} {\mu K} \right)^{n}   \frac{\Gamma\left( \frac{1}{2} + n \right) } {\Gamma\left( \frac{1}{2} \right) }.
$$
This permits us to represent Eq. (\ref{eq:r2n}) as 
\be
\langle r^{2n} \rangle   = \frac{\Gamma \left(\frac{1}{2}\right) \Gamma \left( n  +  \frac{d}{2} \right)}{\Gamma \left(\frac{d}{2}\right) \Gamma \left(n+\frac{1}{2}\right)}  \langle x^{2n} \rangle, 
\label{eq:s2n-z2n}
\ee
which establishes a relation between $\langle x^{2n}\rangle$ and $\langle r^{2n}\rangle$.


The relation in Eq. (\ref{eq:s2n-z2n}) was derived by considering the equilibrium distribution.  
We next verify that the same relation applies for RTP particles.  Since we know that a stationary 
distribution of RTP particles in an isotropic harmonic potential in 2D is \cite{Frydel22c}
\be
p(s) = \frac{\alpha}{\pi}  (1-s^2)^{\alpha-1}, 
\label{eq:pz-2d}
\ee
where we introduce a dimensionless variable $s = \mu K r / v_0$, we can calculate 
the moments $\langle s^{2n} \rangle = \int_0^{1} ds\, 2\pi s p(s) s^{2n}$ which can then compared with 
the moments in Eq. (\ref{eq:z2n-2dl}) for a linear harmonic potential.   The comparison recovers the 
formula in Eq. (\ref{eq:s2n-z2n}) for $d=2$. 

The verification of Eq. (\ref{eq:s2n-z2n}) for $d=3$ is more intricate and the details are relegated to Appendix.  
But it leads to the following relation 
\be
\langle  r^{2n}  \rangle   =   (2n+1) \langle  x^{2n}  \rangle
\label{eq:s2n-a}
\ee
which also agrees with Eq. (\ref{eq:s2n-z2n}) for $d=3$.  Consequently, the relation in Eq. (\ref{eq:s2n-z2n})
is general and applies to passive and RTP particles in a harmonic potential.

Combining the relation in Eq. (\ref{eq:s2n-a}) with the recurrence relation in Eq. (\ref{eq:z2n-3dl}), we next
get the recurrence relation for the moments of a stationary distribution $p(s)$ in an isotropic harmonic 
potential in 3D
\be
\langle s^{2n} \rangle  =  \frac{\alpha}{2n} \sum _{k=0}^{n-1}     \frac{ \langle s^{2k}\rangle}{2n-2k+1}  \frac{(2 k+2)_{\alpha-2}}{(2 n+2)_{\alpha-2}}.  
\label{eq:z2n-3da}
\ee

The central result of this section is Eq. (\ref{eq:s2n-z2n}), which establishes a relation between moments
$\langle r^{2n}\rangle$ and $\langle x^{2n}\rangle$.  This relation is then used to determine 
the recurrence relation in Eq. (\ref{eq:z2n-3da}).

\subsection{recovering $p$ from moments}

To recover a distribution $p(s)$ for an isotropic harmonic trap in 3D from the moments in Eq. (\ref{eq:z2n-3da}), 
we are going to use the Fourier-Legendre expansion, as was done in Sec. (\ref{sec:rec-p3dl}).  
However, since the normalized function is $4\pi s^2 p(s)$, we expand this quantity rather
than $p(s)$.  The resulting expansion is 
\be
4\pi s^2 p(s)  =  2 \sum_{n=0}^{\infty} a_n P_{2n} (s). 
\label{eq:pz-Pn-3d}
\ee
The factor $2$ in front of the sum comes from the fact that $p(s)$ is defined on $[0,1]$ 
while the polynomials $P_n$ are defined on $[-1,1]$.  

The coefficients $a_n$ in the expansion are the same as 
those in Eq. (\ref{eq:an}) but defined in terms of $\langle s^{2n}\rangle$:    
\be
a_n = \frac{4 n+1}{2}  \left[ 2^{2 n} \sum _{k=0}^n \frac{  \langle s^{2 k} \rangle \,  \Gamma \left(k+n+\frac{1}{2}\right) } { (2 k)! (2 n-2 k)! \,  \Gamma \left(k-n+\frac{1}{2}\right)} \right].  
\label{eq:an3d}
\ee

\begin{figure}[hhhh] 
 \begin{center}
 \begin{tabular}{rrrr}
\hspace{-0.3cm}\includegraphics[height=0.16\textwidth,width=0.17\textwidth]{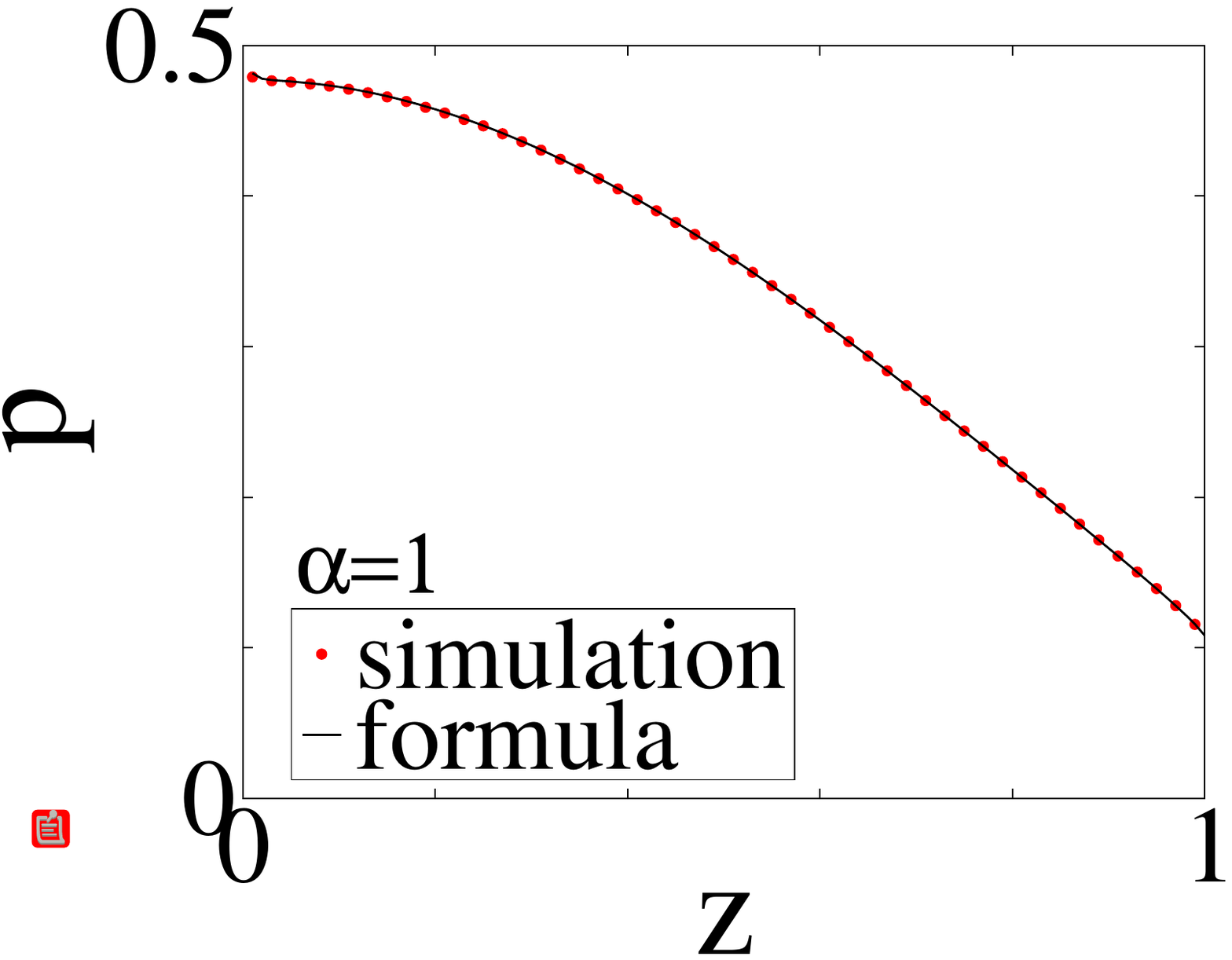} &
\hspace{-0.2cm}\includegraphics[height=0.16\textwidth,width=0.17\textwidth]{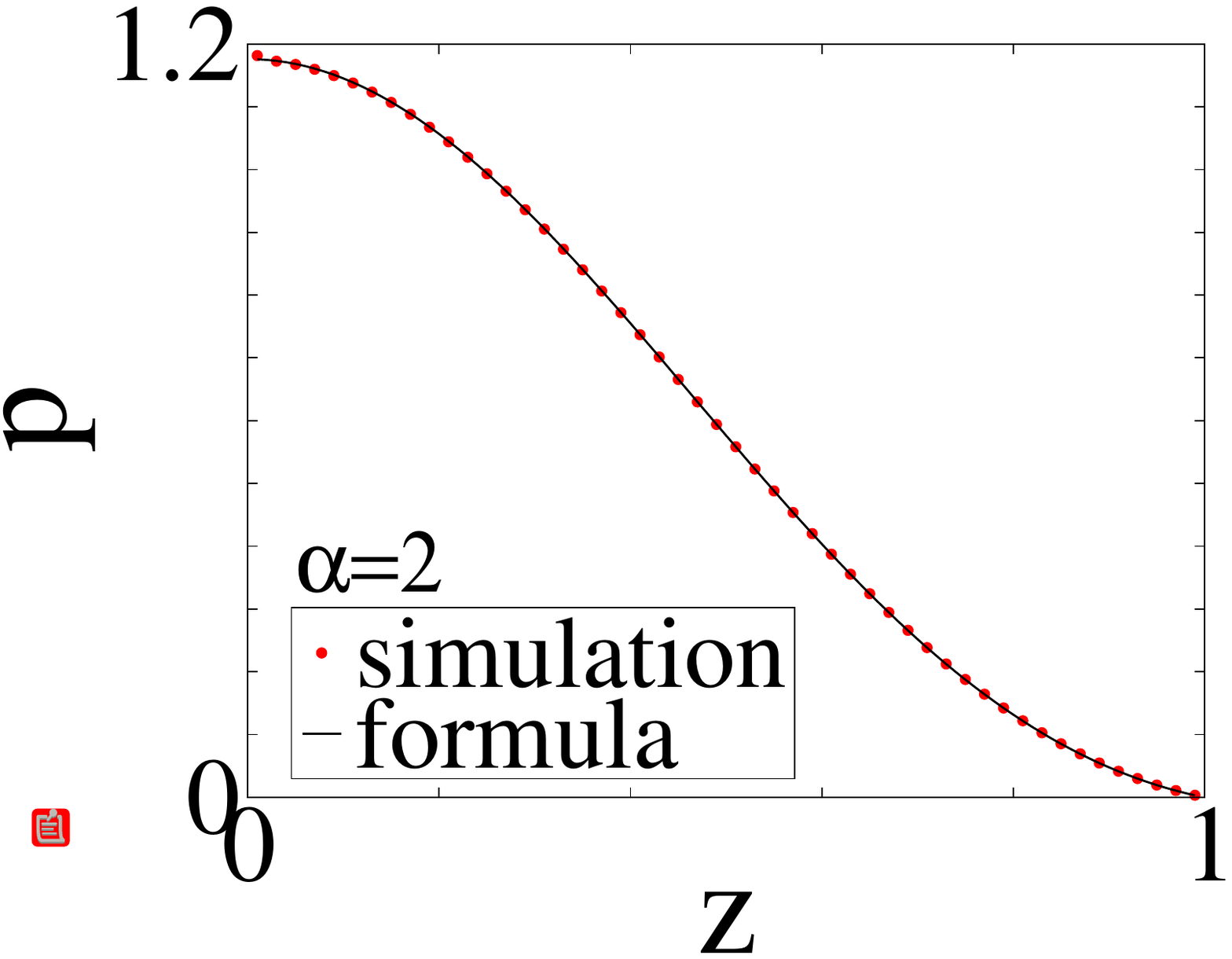} &
\hspace{-0.2cm}\includegraphics[height=0.16\textwidth,width=0.17\textwidth]{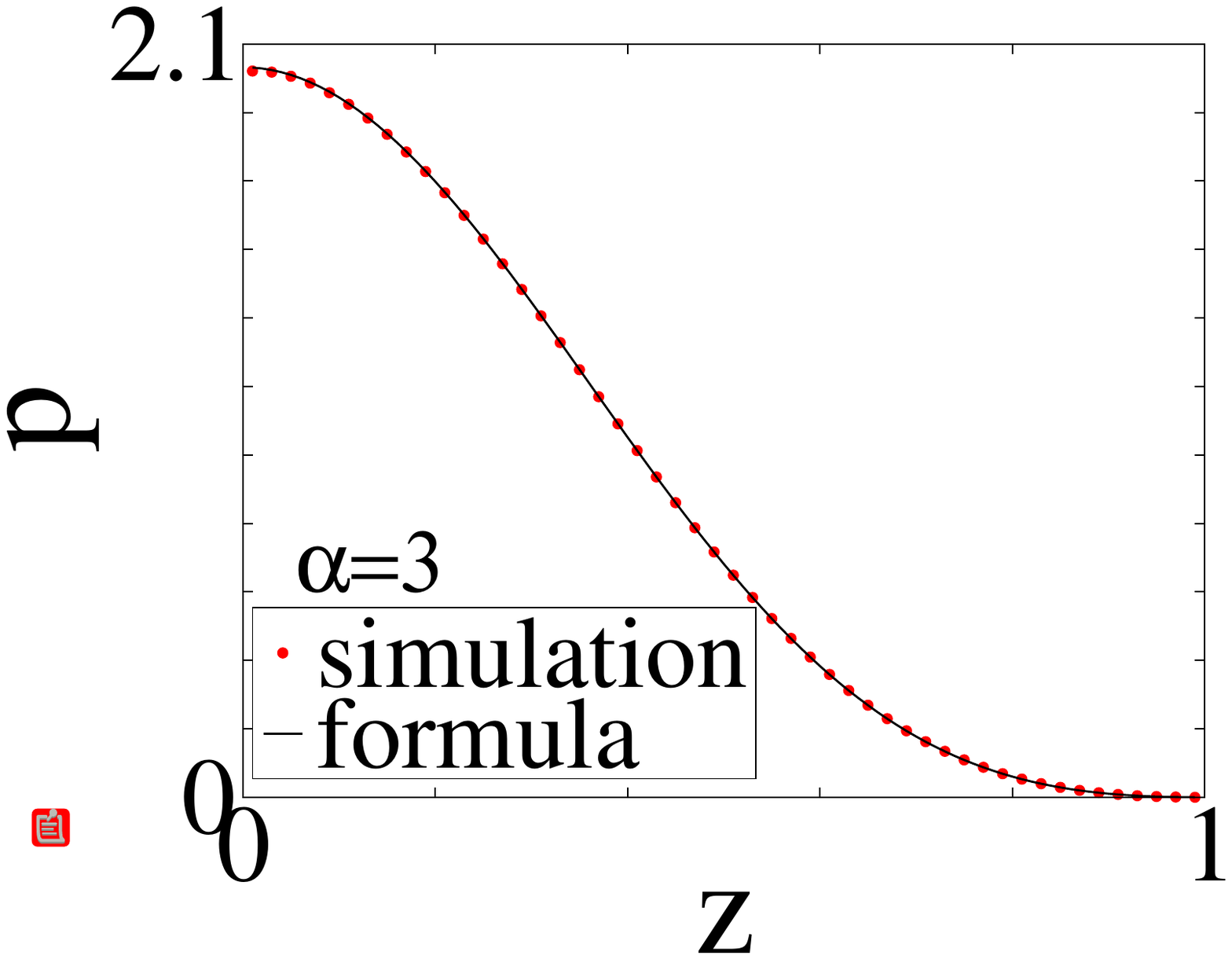} 
 \end{tabular}
 \end{center} 
\caption{Distributions $p$ calculated using the truncated Fourier-Legendre series $p(s) =  \frac{1}{2\pi s^2} \sum_{n=0}^{N_c} a_n P_{2n} (s)$, 
for $N_c=10$ and $a_n$ defined in Eq. (\ref{eq:an3d}). }
\label{fig:fig10} 
\end{figure}

Fig. (\ref{fig:fig10}) compares distributions $p(s)$ obtained using the truncated Fourier-Legendre expansion 
with those obtained from a simulation.  
\graphicspath{{figures/}}
The plots indicate that for $\alpha>1$ the distributions $p(s)$ vanish at $s=1$, confirming $\alpha=1$ to be a point of crossover.

\section{Summary and conclusion}
\label{sec:summary}

The central result of this work is the recurrence relation for generating moments of a stationary distribution 
$p$ for RTP particles in a harmonic trap in three-dimensions.  As there is no available exact expression for $p$ 
in this dimension, this approach provides an alternative route with analytical tractability.

For the potential $u=\frac{1}{2}Kx^2$ the recurrence relation in dimensionless parameters is given in 
Eq. (\ref{eq:z2n-3dl}).  This result is specific for a system embedded in 3D space but it can be 
generalized to any dimension.  A generalized form, valid for any dimension $d$, is given by 
\be
\langle z^{2n} \rangle  =  \frac{\alpha}{2n} \sum _{k=0}^{n-1}     \langle z^{2k}\rangle  \frac{(2 k+1)_{\alpha-1}}{(2 n+1)_{\alpha-1}}  c_{2n-2k}, 
\label{eq:s1}
\ee
where $c_{2n}=A_{0,2n}$ and is given by 
\be
c_{2n}  =  \frac{ \Gamma \left(\frac{d}{2}\right) \Gamma \left(n+\frac{1}{2}\right)}{  \Gamma \left(\frac{1}{2}\right) \Gamma \left(n + \frac{d}{2} \right)}.  
\ee
Note that the parameter of dimensionality only enters via the coefficient $c_{2n}$.

The general formula in Eq. (\ref{eq:s1}) can be verified.  For $d=3$, it recovers the result in Eq. (\ref{eq:z2n-3dl}).  
For $d=1$ and $d=2$, Eq. (\ref{eq:s1}) can be solved for $\langle z^{2n} \rangle$, with solutions found to agree
with Eq. (\ref{eq:z2n-1d}) and Eq. (\ref{eq:z2n-2dl}).      

Using the relation in Eq. (\ref{eq:s2n-z2n}), we can obtain a similar general recurrence relation for the moments of a 
stationary distribution for an isotropic harmonic potential: 
\be
\langle s^{2n} \rangle  =  \frac{\alpha}{2n} \sum _{k=0}^{n-1}    \langle s^{2k}\rangle  \frac{(2 k+2)_{\alpha-2}}{(2 n+2)_{\alpha-2}}   c_{2n-2k}.  
\label{eq:s2}
\ee

The general recurrence formulas in Eq. (\ref{eq:s1}) and in Eq. (\ref{eq:s2}) permit us to better understand 
the role of system dimension.  The fact that in $d=3$ the recurrence relation cannot be solved implies a more 
complex functional form of $p(z)$.  This might help to explain why in this dimension there is no corresponding 
trivial differential equation for $p(z)$ \cite{Frydel22c}.  
The idea of a function without a corresponding differential equation 
was first put forward in 1887 by H\"older \cite{Holder87} who 
considered an Euler gamma function.  In 1900 Hilbert conjectured that the Riemann zeta function
is another example \cite{Hilbert}.  In 1920's this conjecture was proven to be correct \cite{Ostrowski}, and 
in 2015, it was shown that the Riemann zeta function formally satisfies an infinite order linear differential equation \cite{Gorder}.  

{Another important aspect of this work is the determination of a crossover at $\alpha=1$, regardless of a system 
dimension.  The importance of a crossover is that it indicates when to expect the presence of "nearly immobile" 
particles accumulated near a trap border.  Since $\alpha= \frac{1}{\tau \mu K}$ is the ratio of two time scales, 
the persistence time $\tau$ during which an active particle retains its orientation, and the time a particle needs 
to reach a trap border $1/\mu K$, the crossover value gives a way to predict the shape of a distribution once
we know $\alpha$.  If a typical 
persistence time for a {\sl E. Coli} is $\tau \sim 1s$ and a typical velocity $v_0 = 40 \mu s^{-1}$ \cite{book04},   
then we should expect the trap size to be $v_0/\mu K  \lesssim 40 \mu$ in order to see 
accumulation of particles at a trap border.  
}

The most obvious extension of the "recurrence" method is to apply it to other types of active particles, for example, 
the ABP model.  Since the Fokker-Planck equation for the ABP system is different, one expects a different recurrence 
relation.  It is not clear if the methodology is extendeable to other types of external potentials or 
simple interactive systems such as the Kuramoto model \cite{Frydel21}, known to undergoe 
a phase transition, or the one-dimensional asymmetric exclusion process mode \cite{PRL06,
PRE21}. 

\begin{acknowledgments}
D.F. acknowledges financial support from FONDECYT through grant number 1201192.  
\end{acknowledgments}

\section{DATA AVAILABILITY}
The data that support the findings of this study are available from the corresponding author upon 
reasonable request.

\appendix

\section{exact distributions $\rho$ for RTP particles in a harmonic trap in 2D}
\label{sec:app0}

In this section we solve the Fokker-Planck equation in Eq. (\ref{eq:FP-LH}) for $\rho$ by substituting 
for $p$ the expression in Eq. (\ref{eq:pz-2dl}).  
This permits us to posit Eq. (\ref{eq:FP-LH}) as a first order inhomogeneous differential equation:  
\be
0 =   (z-\cos\theta) \rho'   +    (1 -  \alpha) \rho    +    \frac{\alpha A}{2\pi} (1-z^2)^{\alpha - \frac{1}{2}}.  
\label{eq:FP2DL-b}
\ee

By multiplying the above equation by the integrating factor 
$$
e^{\int_{-1}^z dy\,  \frac{1 -  \alpha}{y - \cos\theta} } = \left(  \frac  {\cos\theta + 1}  {\cos\theta-z}    \right)^{\alpha-1}, 
$$
the solution can be represented as 
\be
 \rho_L     =      A\frac{\alpha}{2\pi}  \left | \cos\theta - z   \right | ^{\alpha-1} 
   \int_{-1}^z dz'\,  (1-z'^2)^{\alpha - \frac{1}{2}}    \left | \cos\theta - z'  \right |^{-\alpha}, 
\label{eq:rho2DL-a}
\ee
for the domain $-1<z<\cos\theta$, and 
\be
 \rho_R     =     A \frac{\alpha}{2\pi}  \left |\cos\theta - z   \right| ^{\alpha-1} 
   \int_{z}^1 dz'\,  (1-z'^2)^{\alpha - \frac{1}{2}}    \left | \cos\theta - z'  \right |^{-\alpha},
\label{eq:rho2DL-b}
\ee
for the domain $\cos\theta < z < 1$.

The difference between the solutions in each domain lies in the limits of an integral.  The limits ensure that 
the distribution in Eq. (\ref{eq:rho2DL-a}) vanishes at $z=-1$ and the distribution in Eq. (\ref{eq:rho2DL-b}) 
vanishes at $z=1$.  
Except for $\rho(z,0)$ and $\rho(z,\pi)$, for any other orientation $\theta$, $\rho$ vanishes at both $z=\pm 1$.   
Divergence in $\rho$ comes from the pre-factor $\left | \cos\theta - z   \right | ^{\alpha-1}$.  This implies that 
the divergence for each $\rho$ is localized at $z=\cos\theta$ and the 
crossover corresponds to $\alpha=1$ --- as verified by the behavior of $p_w$.  

\begin{figure}[hhhh] 
 \begin{center}
 \begin{tabular}{rrrr}
\includegraphics[height=0.17\textwidth,width=0.23\textwidth]{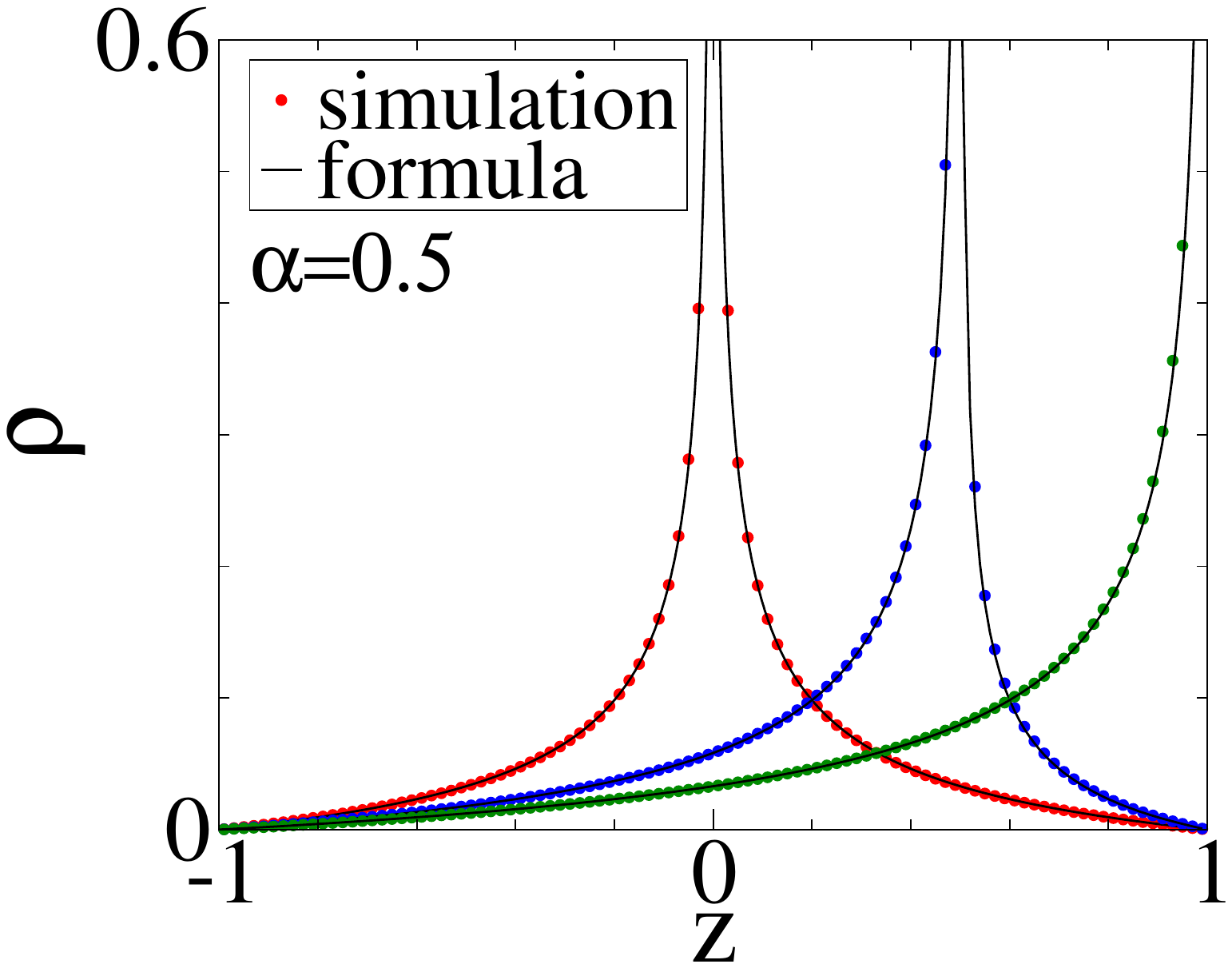} &&
\includegraphics[height=0.17\textwidth,width=0.23\textwidth]{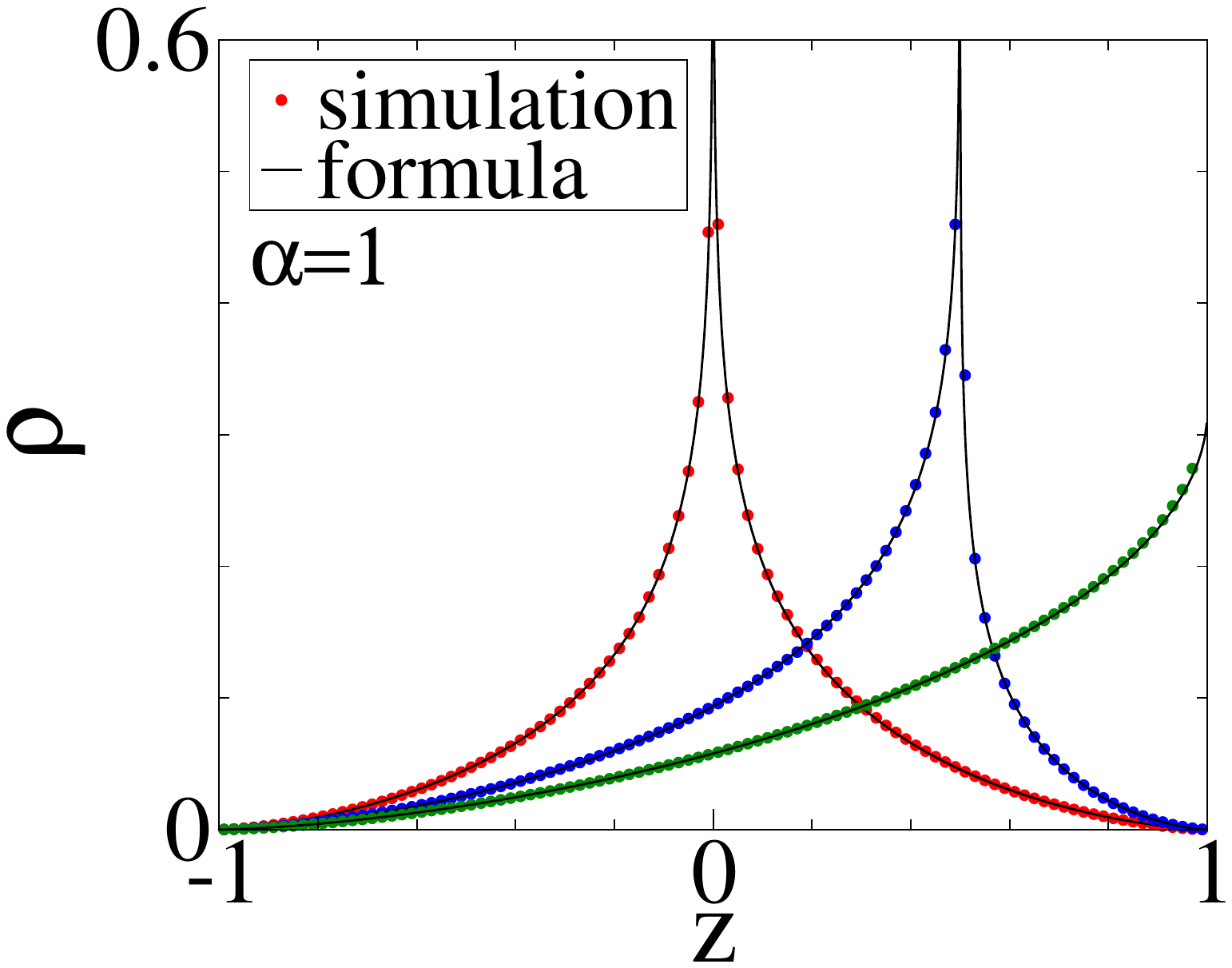}  \\
\includegraphics[height=0.17\textwidth,width=0.23\textwidth]{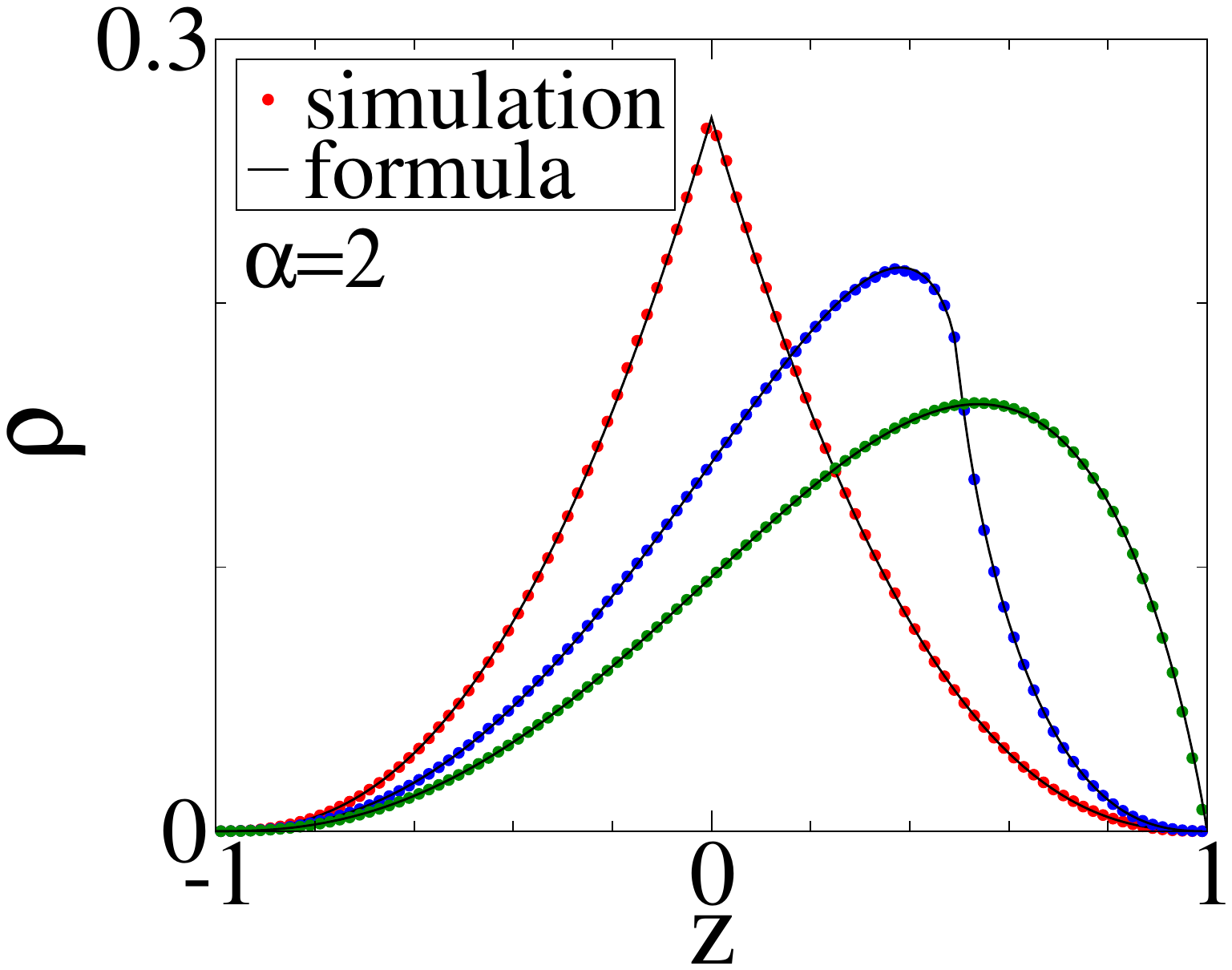}  &&
\includegraphics[height=0.17\textwidth,width=0.23\textwidth]{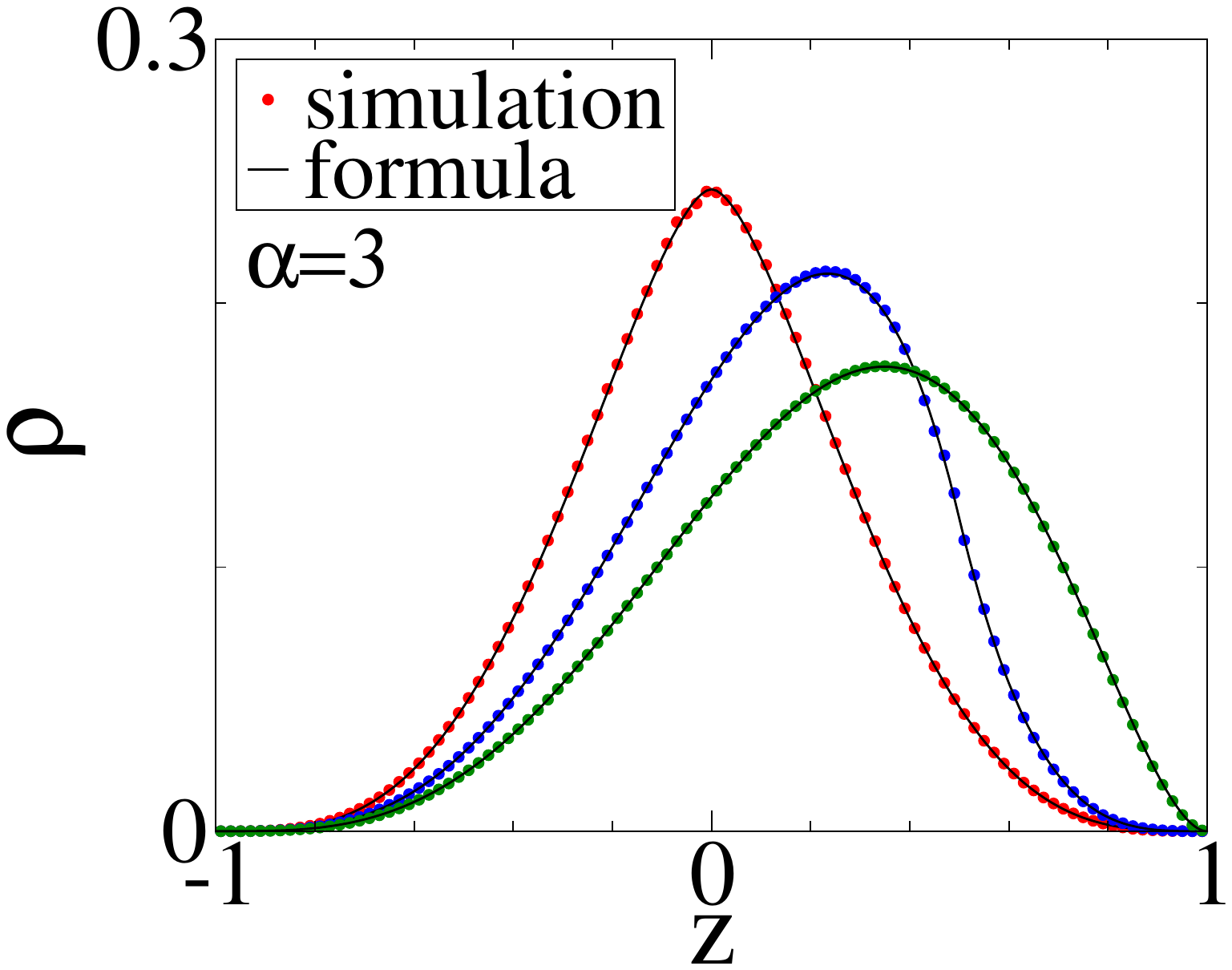}  
 \end{tabular}
 \end{center} 
\caption{ Distributions $\rho(z,\theta)$ for three swimming orientations: $\cos\theta=0,\frac{1}{2},1$ 
--- red, blue , green points, respectively.  
Each distribution integrates to $\int_{-1}^{1}dz\, \rho(z,\theta) = \frac{1}{2\pi}$.  
All circular symbols represent simulation data points, and the lines represent exact results
obtained using Eq. (\ref{eq:rho2DL-a}) and Eq. (\ref{eq:rho2DL-b}).}
\label{fig:fig2} 
\end{figure}

The height of the distribution at $z=\cos\theta$ for $\alpha>1$, when a divergence disappears, can easily be 
evaluated from Eq. (\ref{eq:FP2DL-b}):
\be
\rho(z=\cos\theta,\theta) = \frac{A}{2\pi} \frac{\alpha}{\alpha-1} (1-\cos^2\theta)^{\alpha - \frac{1}{2}}.   
\ee
Note  that the point $z=\cos\theta$ does not represent a maximal value of $\rho$ for $\alpha>1$.  The 
actual peak is shifted toward $z=0$ as shown in Fig. (\ref{fig:fig2}).  
\graphicspath{{figures/}}

The solutions in Eq. (\ref{eq:rho2DL-a}) and Eq. (\ref{eq:rho2DL-b}) 
for $\cos\theta = 0$ become  
\be
\rho  \propto |z|^{\alpha-1} \left(1-z^2\right)^{ \alpha + \frac{1}{2}}
\, _2F_1\left( \frac{\alpha+1}{2}, \frac{2 \alpha+1}{2},\frac{2 \alpha+3}{2}, 1-z^2\right), 
\label{eq:rho0}
\ee
and for $\cos\theta = \pm 1$ 
\be
\rho  \propto \left(1-z^2\right)^{\alpha-1}  (1 \pm z)^{\frac{3}{2}} \, _2F_1 \left(\frac{1}{2}, \frac{2 \alpha + 1}{2} , \frac{2 \alpha + 3}{2}, \frac{1 \pm z}{2}\right).  
\label{eq:rho1}
\ee
Both solutions are for the full domain $[-1,1]$.

It is interesting to compare the distributions given above with the distributions for the three-state RTP 
model in \cite{Basu20}.  The three-state RTP model is an extension of the RTP model in 1D considered 
in Sec. (\ref{sec:sec1}) that includes the zero swimming velocity, $v_{swim} = -v_0,0,v_0$.  
The resulting stationary distribution $p$ has three divergences at $z=-1,0,1$.  The divergences at 
different position correspond to different swimming velocities, for example, 
the divergence at $z=0$ is linked to particles with zero velocity.  The exact solutions for $p_{\pm}$ and $p_0$
are expressed in terms of the hypergeometric functions, similar to the solutions in Eq. (\ref{eq:rho0}) and 
Eq. (\ref{eq:rho1}).  This suggests that the analytical complexity quickly rises if we move away from the 
two-state model.

In Sec. (\ref{sec:sec2A}) we calculated distributions in $w$-space for all particles, that is, averaged over
all orientations.  But knowing $\rho$, it is now possible to calculate distributions in $w$-space corresponding to 
a given orientation.  To obtain such distributions, we transform $\rho$ using the change of variables 
$z = -w + \cos\theta$. 

\begin{figure}[hhhh] 
 \begin{center}
 \begin{tabular}{rrrr}
\includegraphics[height=0.17\textwidth,width=0.23\textwidth]{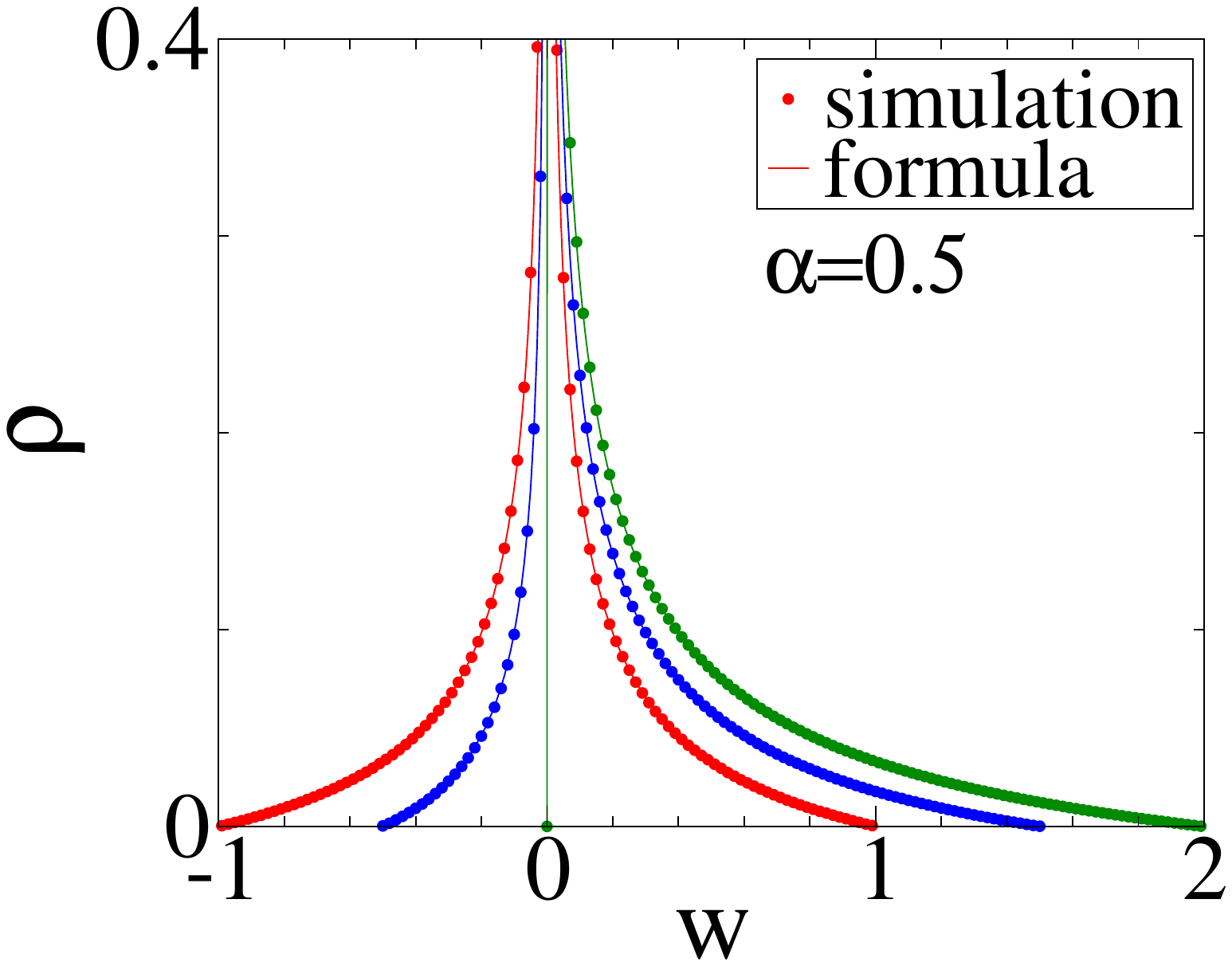} &&
\includegraphics[height=0.17\textwidth,width=0.23\textwidth]{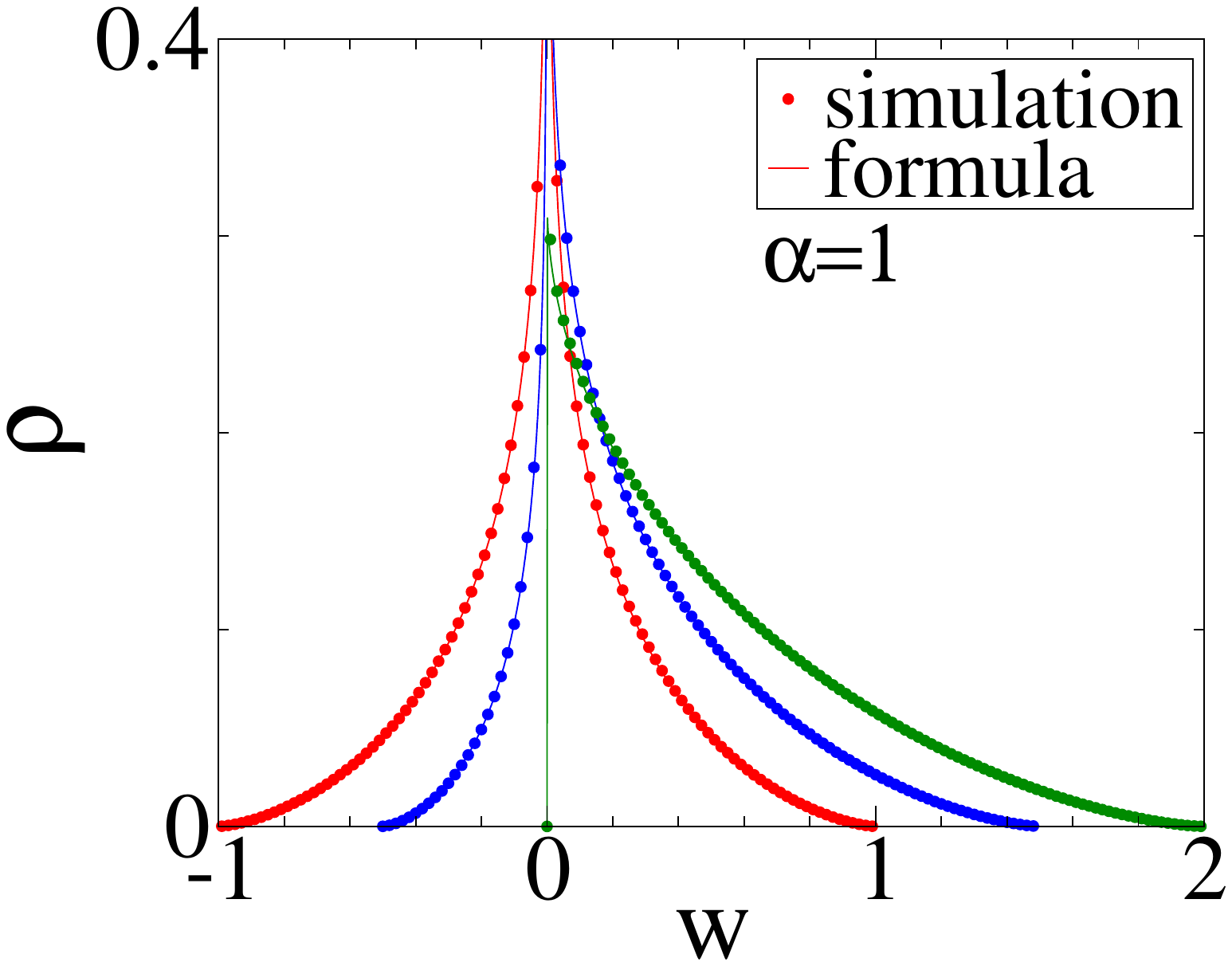}  \\
\includegraphics[height=0.17\textwidth,width=0.23\textwidth]{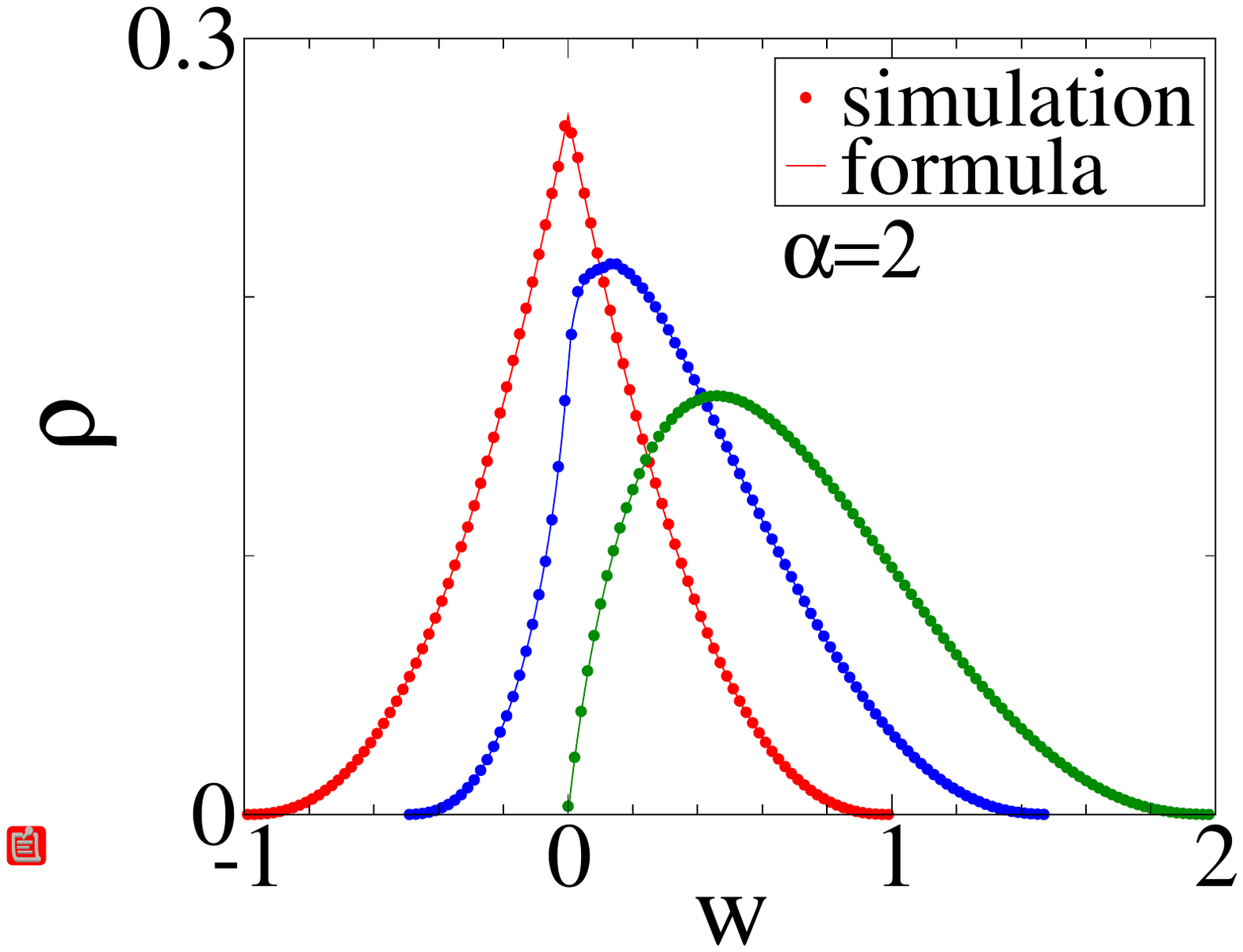}  &&
\includegraphics[height=0.17\textwidth,width=0.23\textwidth]{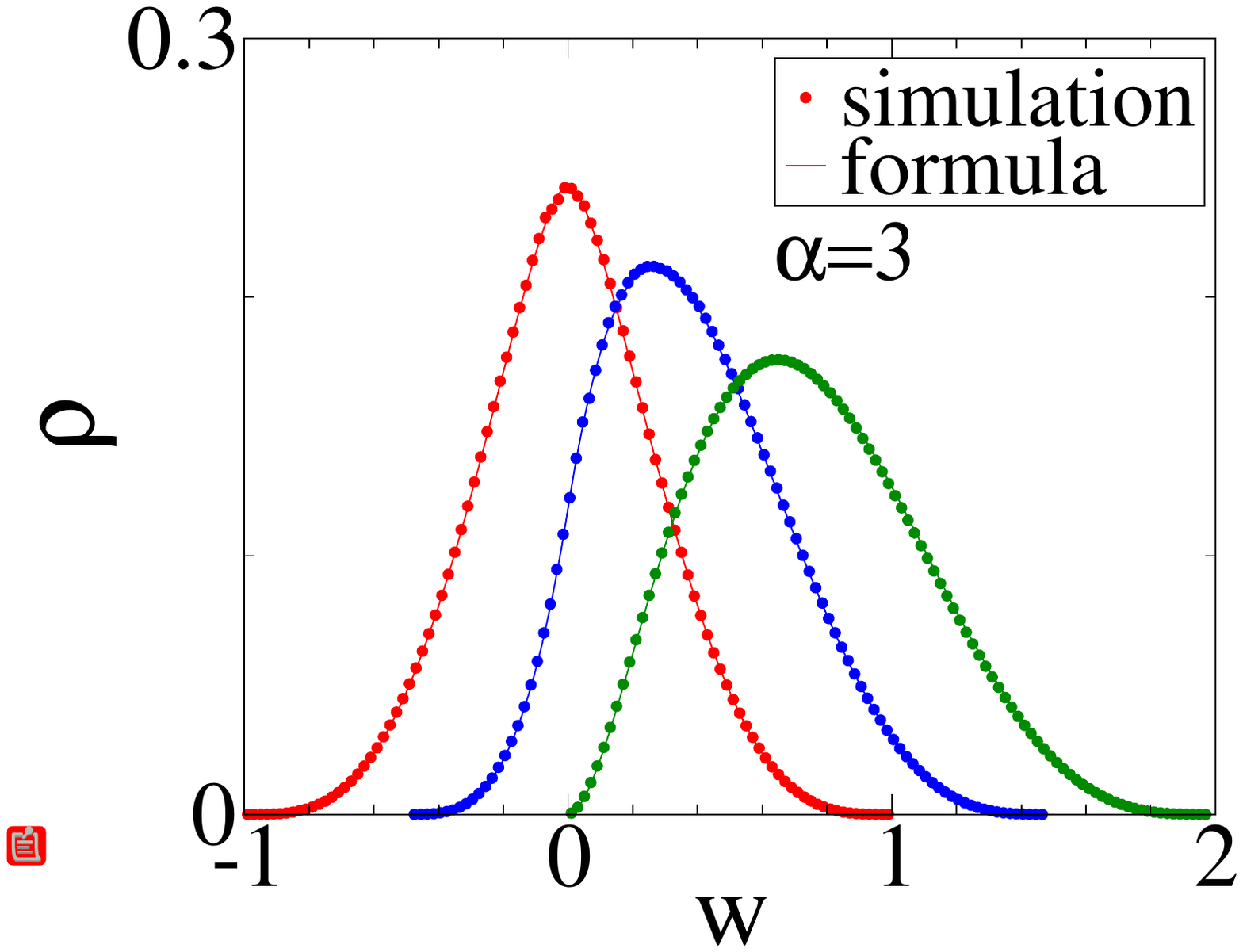}  
 \end{tabular}
 \end{center} 
\caption{ Distributions $\rho(w,\theta)$ for three swimming orientations: $\cos\theta=0,\frac{1}{2},1$
--- red, blue , green points, respectively.  
All circular symbols represent simulation data points, and the lines represent exact results
obtained using Eq. (\ref{eq:rho2DL-a}) and Eq. (\ref{eq:rho2DL-b}).}
\label{fig:fig3} 
\end{figure}

The distributions $\rho(w,\theta)$ are plotted in Fig. (\ref{fig:fig3}). 
\graphicspath{{figures/}}
Note that all divergences, regardless of the orientation of a motion, are at $w = 0$ which signals the presence 
of nearly immobile particles. For $\alpha > 1$, nearly immobile particles disappear, manifested by the disappearance of 
divergences.   Another observation is that for $\alpha>1$, all the peaks, originally at $w=0$, start to shift away from 
$w = 0$ toward $w\to\cos\theta$, that is, the value of a swimming velocity of particles with orientation $\theta$.  
Only distribution with orientation $\theta=\pm \pi/2$ are centered around $w=0$.  
 
Note that the domain of a distribution $\rho(w,\theta)$ depends on $\cos\theta$ and is given by $w\in (-1 + \cos\theta,1+\cos\theta)$.

\section{exact results for $p_w$}
\label{sec:app1}

\begin{figure}[hhhh] 
 \begin{center}
 \begin{tabular}{rrrr}
\hspace{-0.3cm}\includegraphics[height=0.16\textwidth,width=0.17\textwidth]{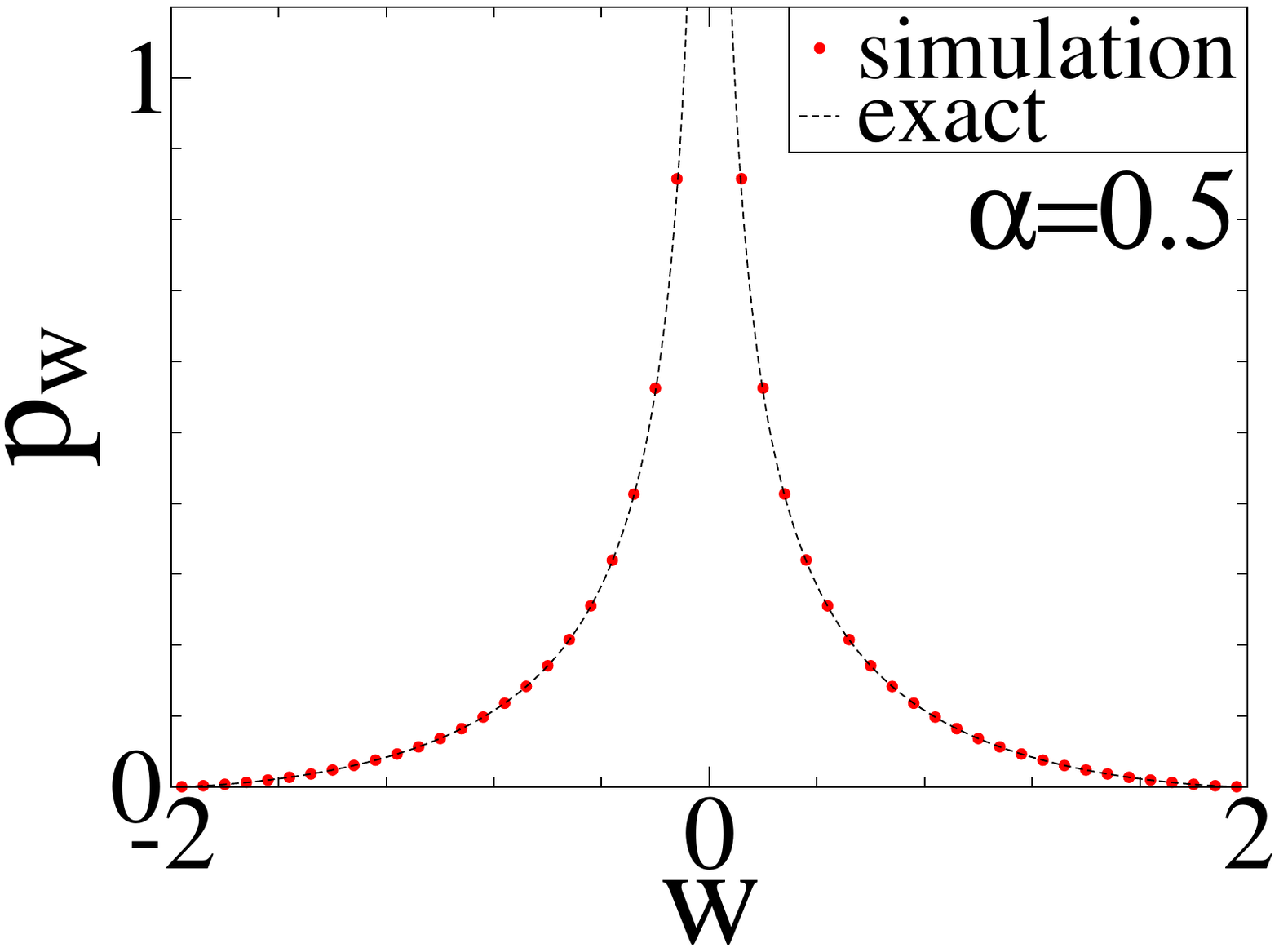} &
\hspace{-0.2cm}\includegraphics[height=0.16\textwidth,width=0.17\textwidth]{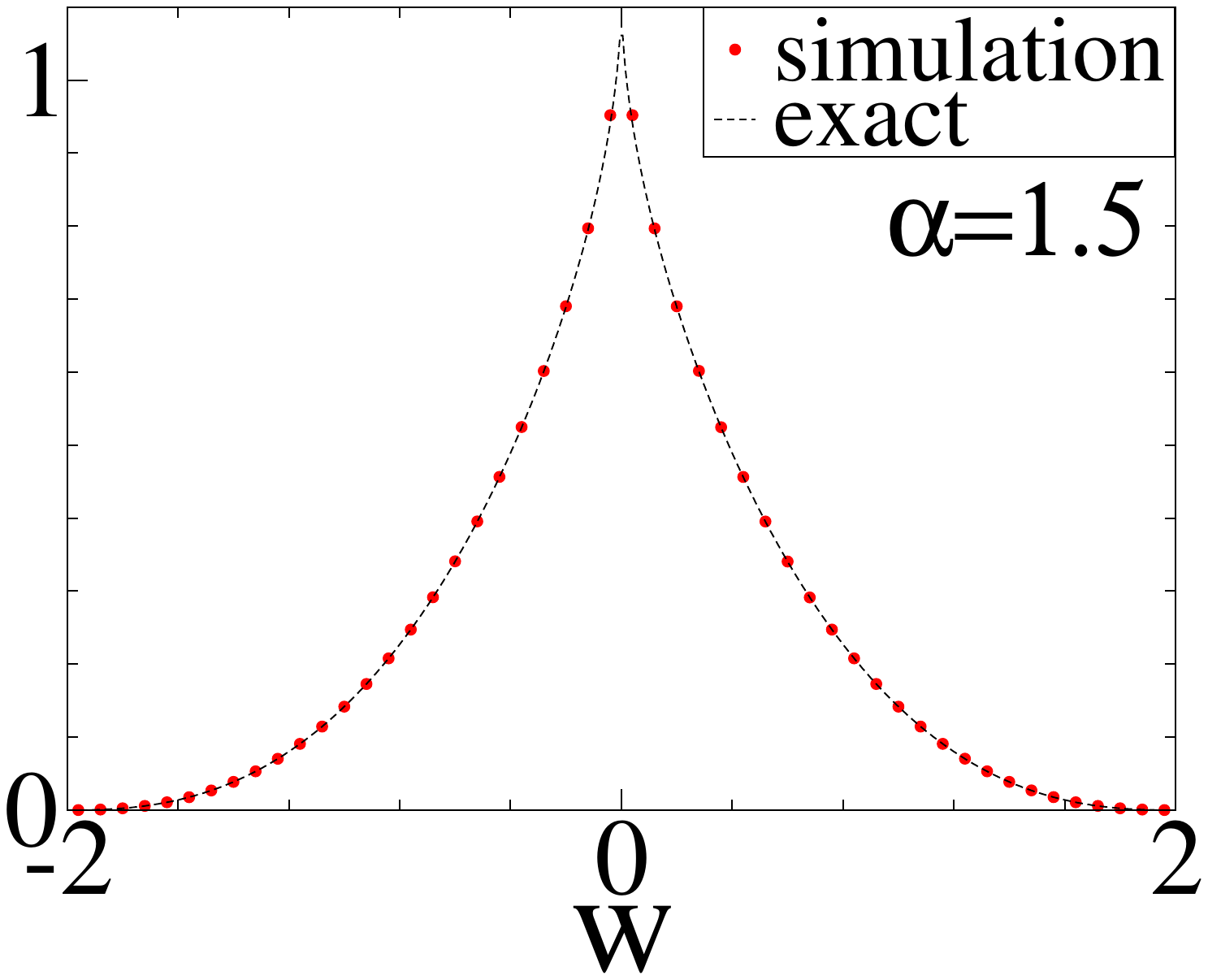} &
\hspace{-0.2cm}\includegraphics[height=0.16\textwidth,width=0.17\textwidth]{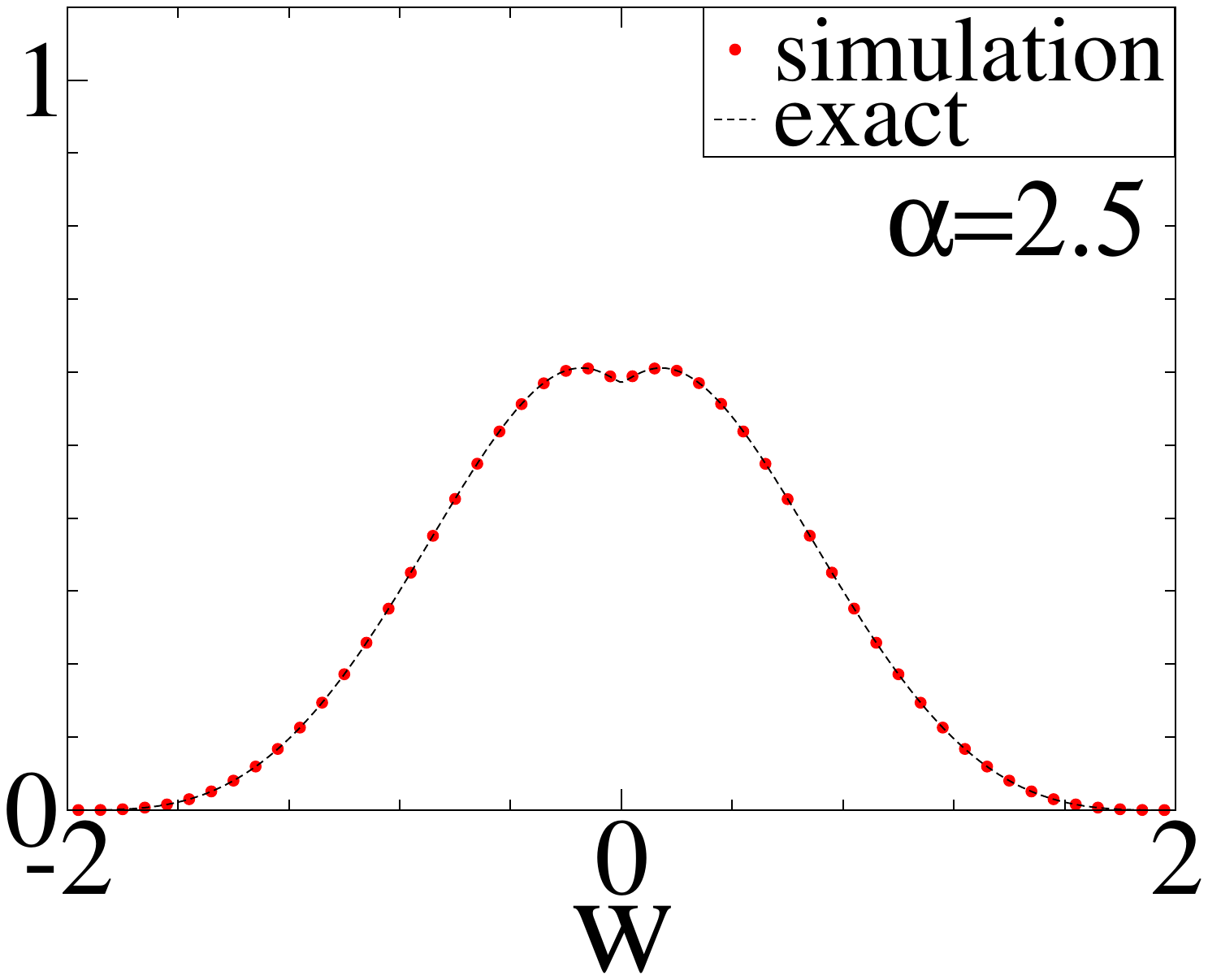} 
 \end{tabular}
 \end{center} 
\caption{ Distributions $p_w$ for different values of $\alpha$.    
Exact distributions (represented by lines) are obtained from Eq. (\ref{eq:pw-exact}).  The circular symbols 
represent simulation data points.}
\label{fig:fig5} 
\end{figure}

For half integer values of $\alpha$, the integral in Eq. (\ref{eq:pw-2dl}) can be evaluated exactly.  
For the first three values, $\alpha=\frac{1}{2},\frac{3}{2},\frac{5}{2}$, the distribution can be represented as 
\be
p_w = \frac{a_{\alpha}}{\pi} \sqrt{\frac{2-w}{w}}   +   b_{\alpha}  \left[ \frac{\sin ^{-1}(1-w)}{2 \pi } + \frac{1}{4} \right], 
\label{eq:pw-exact}
\ee
where $a_{\alpha}$ and $b_{\alpha}$ are polynomials given in 
Table (\ref{table1}).  
\begin{table}[h!]
\centering
 \begin{tabular}{ l | l l } 
 \hline
 & \\[-1.8ex]
$\alpha$ &  ~~~ $a_{\alpha}$  &  $b_{\alpha}$ \\ [0.2ex] 
 \hline
 & \\[-1ex]
${1}/{2}$ &   ~~~  {\scriptsize $1$ }   &   {\scriptsize $-1$ } \\ [1.ex] 
${3}/{2}$ &   ~~~  {\scriptsize $- \left(7w + 13 w^2 \right)/8$}    &   {\scriptsize $\left(9 + 6 w^2 \right) / 4 $ }\\ [1.ex] 
${5}/{2}$ &   ~~~  {\scriptsize $  \left(75w + 537 w^2 + 178 w^3 + 298 w^4 \right) / 128 $}    &   {\scriptsize $ \left(75  -  600 w^2  - 120w^4\right) / 64$ } \\ [1.ex] 
  \hline
\end{tabular}
\caption{Polynomials $a_{\alpha}$ and $b_{\alpha}$ for the formula in Eq. (\ref{eq:pw-exact}).}
\label{table1}
\end{table}
Fig. (\ref{fig:fig5}) compares these analytical results with simulation data points.  
\graphicspath{{figures/}}

\section{Convolution of probability distributions}
\label{sec:app2}


To show explicitly that the convolution of probability distributions works for particles in a harmonic potential $u=Kr^2/2$
whose dynamics is governed by $N$ independent forces ${\bf f}_i$, we 
consider the Langevin equation.  For an unconfined system the Langevin equation is 
\be
\dot {\bf r} =  \mu \sum_{i=1}^N {\bf f}_i, 
\ee
where 
$
\langle {\bf f}_i(t) \cdot{\bf f}_j(t') \rangle  =  f_i^2 \delta_{ij} \, \tau_i^{-1} e^{-|t-t'|/\tau_i},
$
where we assumed exponential memory such that in the limit $\tau_i\to 0$, 
$
\tau_i^{-1}  e^{-|t-t'|/\tau_i} \to \delta(t-t').  
$
For particles in a harmonic trap the Langevin equation is 
\be
\dot {\bf r} =  \mu \sum_{i=1}^N {\bf f}_i   -  \mu K  {\bf r},
\label{eq:L}
\ee
and can be solved to yield 
\be
{\bf r}    =      \mu  \sum_{i=1}^N \int_{-\infty}^t ds\,   e^{ -\mu K (t-s )}  {\bf f}_i(s).
\ee
Using the above solution, the Langevin equation in Eq. (\ref{eq:L}) can be represented as 
\be
\dot {\bf r} =  \mu \sum_{i=1}^N \bar {\bf f}_i   
\ee
where the new forces are defined as 
\be
\bar {\bf f}_i =  {\bf f}_i   -   \mu K \int_{-\infty}^t ds\,   e^{ -\mu K (t-s )} {\bf f}_i(s). 
\ee
The new forces can be shows to be independent:  
\be
\langle \bar {\bf f}_i(t) \cdot \bar {\bf f}_j(t') \rangle    =   f_i^2 \delta_{ij} M(t-t'),
\ee
where $M$ is a resulting new memory function.  


%

\section{computation of moments for an isotropic harmonic potential in 3D}
\label{sec:app3}

In this section we provide an explicit verification of the relation in Eq. (\ref{eq:s2n-z2n}) 
for the RTP particles in a harmonic trap in 3D.  
We start with the stationary FP equation 
$$
0  =  \bnabla \cdot \left[ \left( \mu K {\bf r} - v_0 {\bf n} \right) \rho \right]   +  \hat L\, \rho, 
$$
where the operator $\hat L$ for the RTP model in 3D is given by 
$$
\hat L\, \rho = \frac{1}{\tau} \left[ -\rho + \frac{1}{2} \int_0^{\pi} d\theta\, \sin\theta \rho(x,\theta) \right].  
$$
Since $\rho$ depends on the relative orientation of the vectors ${\bf r}$ and ${\bf n}$, we fix 
${\bf n}$ along the $x$-axis, ${\bf n} = {\bf e}_x$, and the FP equation becomes 
\be
0 =  \mu K  {\bf r}  \cdot \bnabla \rho  +  \mu K \rho (\bnabla \cdot {\bf r}) - v_0  \frac{\partial \rho}{\partial x}  
+  \hat L\, \rho.  
\ee
In spherical coordinates: 
$\frac{\partial \rho}{\partial x} = \cos\theta \frac{\partial\rho}{\partial r} -   \frac{\sin\theta}{r} \frac{\partial \rho}{\partial \theta}$,
${\bf r} \cdot \bnabla \rho  = r \frac{\partial \rho}{\partial r}$, and $\bnabla \cdot {\bf r} = d$.  
The FP equation in reduced units $s = \mu K r/v_0$ can now be expressed 
\be
0 =   s \rho'  +  3 \rho   -     \cos\theta \rho'  +    \frac{\sin\theta}{s} \frac{\partial \rho}{\partial \theta}   -  \alpha \rho    +    \frac{\alpha}{2} p, 
\ee
where $$p = \int_0^{\pi} d\theta\, \sin\theta \rho(z,\theta).$$  
By multiplying the above equation by $s^l \cos^m\theta$
and then integrating it as $4\pi \int_0^{\infty} ds\, s^2 \int_0^{\pi} d\theta\, \sin\theta$, we get the following 
recurrence relation:  
\be
B_{l,m}   =   \frac{\alpha} { l + \alpha } B_{l,0} B_{0,m}     +    \frac{ l - m }{ l + \alpha } B_{l-1,m+1}   +    \frac{m}{l+\alpha} B_{l-1,m-1}. 
\label{eq:Blm}
\ee
where 
$
B_{l,m} = \langle s^l \cos^m\theta \rangle,
$
and the initial condition is
$
B_{0,0} = 1.  
$
The recurrence relation in Eq. (\ref{eq:Blm}) is used to solve for $B_{2n,0} = \langle s^{2n}\rangle$ that satisfies the relation 
\be
\langle s^{2n} \rangle =  (2n+1) \langle z^{2n} \rangle_{},
\ee
where $\langle z^{2n} \rangle$ is defined in Eq. (\ref{eq:z2n-3dl}).  Consequently, the relation in Eq. (\ref{eq:s2n-z2n}) 
is verified for the RTP model.




\begin{thebibliography}{99}

\bibitem{Tailleur08}
J. Tailleur and M. E. Cates, 
{\sl Statistical Mechanics of Interacting Run-and-Tumble Bacteria}, 
Phys. Rev. Lett. {\bf 100}, 218103 (2008).

\bibitem{Tailleur09}
J. Tailleur and M. E. Cates, 
{\sl Sedimentation, trapping, and rectification of dilute bacteria}, 
Europhys. Lett. {\bf 86}, 60002 (2009).

\bibitem{Basu20}
U. Basu, S. N. Majumdar, A. Rosso, S. Sabhapandit, and G. Schehr, 
{\sl Exact stationary state of a run-and-tumble particle with three internal states in a harmonic trap}, 
J. Phys. A {\bf 53}, 09LT01 (2020).

\bibitem{Dhar19}
A. Dhar, A. Kundu, S. N. Majumdar, S. Sabhapandit, and G. Schehr, 
{\sl Run-and-tumble particle in one-dimensional confining potentials: Steady-state, relaxation, and first-passage properties}, 
Phys. Rev. E {\bf 99}, 032132 (2019).

\bibitem{Frydel22c}
D. Frydel,
{\sl Positing the problem of stationary distributions of active particles as third-order differential equation}, 
Phys. Rev. E {\bf 106}, 024121 (2022). 

\bibitem{Frydel23}
D. Frydel,
{\sl Entropy production of active particles formulated for underdamped dynamics}, 
Phys. Rev. E {\bf 107}, 014604 (2023).


\bibitem{Scher22}
N. R. Smith, P. Le Doussal, S. N. Majumdar, G. Schehr, 
{\sl Exact position distribution of a harmonically confined run-and-tumble particle in two dimensions}, 
Phys. Rev. E {\bf 106}, 054133 (2022).

\bibitem{Dhar20}
K. Malakar, A. Das, A. Kundu, K. V. Kumar, and A. Dhar
{\sl Steady state of an active Brownian particle in a two-dimensional harmonic trap}, 
Phys. Rev. E {\bf 101}, 022610 (2020).


\bibitem{Szamel14}
G. Szamel, 
{\sl Self-propelled particle in an external potential: Existence of an effective temperature},
Phys, Rev. E {\bf 90}, 012111 (2014).



\bibitem{Carpini22}
L. Caprini, A. R. Sprenger, H. L\"owen, R Wittmann, 
{\sl The parental active model: A unifying stochastic description of self-propulsion},  
J. Chem. Phys. {\bf 156}, 071102 (2022).

\bibitem{Cargalio22}
M. Caraglio, T. Franosch, 
{\sl Analytic solution of an active brownian particle in a harmonic well},
Phys. Rev. Lett. {\bf 129}, 158001, (2022).  









\bibitem{Farago22}
N. R. Smith and O. Farago, 
{\sl Nonequilibrium steady state for harmonically confined active particles}, 
Phys. Rev. E {\bf 106}, 054118 (2022).



\bibitem{Gupta21}
D. Gupta and D. A. Sivak, 
{\sl Heat fluctuations in a harmonic chain of active particles}, 
Phys.Rev. E {\bf 104}, 024605 (2021).

\bibitem{Kundu21}
P. Singh and A. Kundu, 
{\sl Crossover behaviours exhibited by fluctuations and correlations in a chain of active particles}, 
J. Phys. A: Math. Theor. {\bf 54}, 305001 (2021).  

\bibitem{Basu22}
I. Santra, U. Basu, 
{\sl Activity driven transport in harmonic chains}, 
SciPost Phys. {\bf 13}, 041 (2022). 


\bibitem{Brady16}
 S. C. Takatori, R. De Dier, J. Vermant, and J. F. Brady, 
 {\sl Acoustic trapping of active matter},
 Nat. Commun. {\bf 7}, 10694 (2016).

\bibitem{Lowen22}
I. Buttinoni, L. Caprini, L. Alvarez , F. J. Schwarzendahl, and H. L\"owen, 
{\sl Active colloids in harmonic optical potentials},
EPL {\bf 140},  27001 (2022). 

\bibitem{Pruessner21}
R. Garcia-Millan and G. Pruessner 
{\sl Run-and-tumble motion in a harmonic potential: field theory and entropy production}, 
J. Stat. Mech. 063203 (2021).  

\bibitem{Frydel22a}
D. Frydel,
{\sl Intuitive view of entropy production of ideal run-and-tumble particles}, 
Phys. Rev. E {\bf 105}, 034113 (2022).




\bibitem{Dabelow19}
L Dabelow, S Bo, R Eichhorn, 
{\sl Irreversibility in active matter systems: Fluctuation theorem and mutual information}, 
Phys. Rev. X {\bf 9}, 021009, (2019).   

\bibitem{Caprini19}
L. Caprini, U. M. B. Marconi, A. Puglisi, A. Vulpiani, 
{\sl The entropy production of Ornstein-Uhlenbeck active particles: a path integral method for correlations}, 
J. Stat. Mech. 053203, (2019).

\bibitem{Dabelow21}
L. Dabelow, S. Bo, R. Eichhorn,
{\sl How irreversible are steady-state trajectories of a trapped active particle?}, 
J. Stat. Mech. 033216, (2021).




\bibitem{Maggi14}
C. Maggi, M. Paoluzzi, N. Pellicciotta, A. Lepore, L. Angelani, R. Di Leonardo, 
{\sl Generalized energy equipartition in harmonic oscillators driven by active baths}, 
Phys. Rev. Lett. {\bf 113}, 238303, (2014).
 
\bibitem{Martin21}
D. Martin, J. O’Byrne, M. E. Cates, E. Fodor, C. Nardini, J. Tailleur, and F. van Wijland, 
{\sl Statistical mechanics of active Ornstein-Uhlenbeck particles}, 
Phys. Rev.  E {\bf 103}, 032607 (2021).  

\bibitem{Caprini21}
L. Caprini, U. M. B. Marconi, 
{\sl Inertial self-propelled particles}, 
J. Chem. Phys. {\bf 154}, 024902 (2021). 














\bibitem{Frydel21b}
D. Frydel, "{\sl Generalized run-and-tumble model for an arbitrary distribution of velocities in 1D geometry}", 
J. Stat. Mech.: Theory Exp. {\bf 184},  083220 (2021). 


\bibitem{Frydel21a}
D. Frydel, 
{\sl Stationary distributions of propelled particles as a system with quenched disorder}, 
Phys. Rev. E {\bf 103}, 052603 (2021).  



%
%
%



%

































%
%
%




\bibitem{book08}
N. F. Sharpe and R. F. Carter, Genetic Testing (John Wiley \& Sons, Inc., Hoboken, NJ, 2005).
Dimitri P. Bertsekas and John N. Tsitsiklis, {\sl Introduction To Probability}
(Athena Scientific 2008) 2nd Ed. 

\bibitem{book19}
Joseph K. Blitzstein and Jessica Hwang, {\sl Introduction to Probability}
 (Chapman \& Hall/CRC Texts in Statistical Science 2019) 2nd Ed.  






\bibitem{Holder87}
O. H\"older, 
{\sl \"Uber die Eigenschaft der $\Gamma$-Function, keiner algebraischen Differentialgleichung zu gen\"ugen}. 
Math Ann {\bf 28}, 1 1887.

\bibitem{Hilbert}
D. Hilbert, 
{\sl Mathematische Probleme, in: Die Hilbertschen Probleme}, 
Akademische Verlagsgesellschadt Geest \& Portig, Leipzig pp. 23-80,  (1971) . 

\bibitem{Ostrowski}
A. Ostrowski, {\sl \"Uber Dirichletsche Reihnen und algebraische Differentialgleichungen}, 
Math Z {\bf 8}, 241 (1920). 

\bibitem{Gorder}
R. A. Van Gorder, 
{\sl Does the Riemann zeta function satisfy a differential equation},
J. Number Theory {\bf 147}, 778 (2015).




\bibitem{book04}
Berg, Howard , {\sl E. coli in motion}, (Springer-Verlag 2004).  


\bibitem{Frydel21}
D. Frydel, {\sl Kuramoto Model with run-and-tumble dynamics}, 
Phys. Rev. E {\bf 104}, 024203 (2021). 







\bibitem{PRL06}
H. Hinsch and E. Frey, 
{\sl Bulk-Driven Nonequilibrium Phase Transitions in a Mesoscopic Ring}, 
Phys. Rev. Lett. {\bf 97}, 095701 (2006). 

%
%

\bibitem{PRE21}
A. Haldar, P. Roy, and A. Basu, 
{\sl Asymmetric exclusion processes with fixed resources: Reservoir crowding and steady states}, 
Phys.Rev. E {\bf 104}, 034106 (2021).





%


\end{thebibliography}
\end{document}